\documentclass[fleqn,usenatbib]{mnras}

\usepackage{newtxtext,newtxmath}
\usepackage[T1]{fontenc}
\usepackage{ae,aecompl}

\usepackage{graphicx}	% Including figure files
\usepackage{amsmath}	% Advanced maths commands
\usepackage{amssymb}	% Extra maths symbols
\usepackage{caption}
\usepackage[flushleft]{threeparttable}

\newcommand{\es}{\,erg~s$^{-1}$~} % erg/s
\newcommand{\kms}{\,km~s$^{-1}$~} % km/s
\newcommand{\fluxden}{\,erg~s$^{-1}$~cm$^{-2}$~} % erg/s/cm2
\newcommand{\flux}{\,erg~s$^{-1}$~cm$^{-2}$~\AA$^{-1}$~} % erg/s/cm2/A

\newcommand{\lya}{\,Ly$\alpha$~}
\newcommand{\oii}{\,[O \textsc{ii}]~}
\newcommand{\oiii}{\,[O \textsc{iii}]~}

\title[Faint radio galaxies at high redshifts]{The nature of faint radio galaxies at high redshifts}

\author[A. Saxena et al.]{A.~Saxena$^{1,2}$\thanks{E-mail: saxena@strw.leidenuniv.nl}, H.~J.~A.~R{\"o}ttgering$^{1}$, K.~J.~Duncan$^1$, G.~J.~Hill$^{3,4}$, P.~N.~Best$^5$,
\newauthor B.~L.~Indahl$^3$, M.~Marinello$^6$, R.~A.~Overzier$^{6,7}$, L.~Pentericci$^2$, I.~Prandoni$^8$,
\newauthor H.~Dannerbauer$^{9,10}$ and R.~Barrena$^{9,10}$
\\ \\
$^{1}$Leiden Observatory, Leiden University, P.O. Box 9513, 2300 RA Leiden, The Netherlands\\
$^{2}$INAF-Osservatorio Astronomico di Roma, Via Frascati 33, I-00040 Monteporzio (RM), Italy\\
$^{3}$Department of Astronomy, University of Texas at Austin, 2515 Speedway, Stop C1400, Austin, TX 78712, USA\\
$^{4}$McDonald Observatory, University of Texas at Austin, 2515 Speedway, Stop C1402, Austin, TX 78712, USA\\
$^{5}$Institute for Astronomy, University of Edinburgh, Royal Observatory, Blackford Hill, Edinburgh EH9 3HJ, UK\\
$^{6}$Observat\'orio Nacional, Rua General Jos\'e Cristino, 77, S\~ao Crist\'ov\~ao, Rio de Janeiro, RJ, CEP 20921-400, Brazil\\
$^{7}$Institute of Astronomy, Geophysics and Atmospheric Sciences, Department of Astronomy, University of S\~ao Paulo, S\~ao Paulo,\\
  SP 05508-090, Brazil\\
$^{8}$INAF-Instituto di Radioastronomia, Via P. Gobetti 101, I-40129 Bologna, Italy\\
$^{9}$Instituto de Astrof\'{i}sica de Canarias (IAC), E-38205 La Laguna, Tenerife, Spain\\
$^{10}$Universidad de La Laguna, Dpto. Astrof\'{i}sica, E-38206 La Laguna, Tenerife, Spain
}

\date{Accepted 2019 September 4. Received 2019 September 4; in original form 2019 June 3}

\pubyear{2019}

\begin{document}
\label{firstpage}
\pagerange{\pageref{firstpage}--\pageref{lastpage}}
\maketitle

\begin{abstract}
We present spectra and near-infrared images of a sample of faint radio sources initially selected as promising high-redshift radio galaxy (HzRG) candidates. We have determined redshifts for a total of 13 radio galaxies with redshifts ranging from $0.52\le z \le 5.72$. Our sample probes radio luminosities that are almost an order of magnitude fainter than previous large samples at the highest redshifts. We use near-infrared photometry for a subsample of these galaxies to calculate stellar masses using simple stellar population models, and find stellar masses to be in the range $10^{10.8} - 10^{11.7} M_\odot$. We then compare our faint radio galaxies with brighter radio galaxies at $z\ge2$ from the literature. We find that fainter radio galaxies have lower Ly$\alpha$ luminosities and narrower line widths compared to the bright ones, implying photoionisation by weaker AGN. We also rule out the presence of strong shocks in faint HzRGs. The stellar masses determined for faint HzRGs are lower than those observed for brighter ones. We find that faint HzRG population in the redshift range $2-4$ forms a bridge between star-forming and narrow-line AGN, whereas the ones at $z>4$ are likely to be dominated by star-formation, and may be building up their stellar mass through cold accretion of gas. Finally, we show that the overall redshift evolution of radio sizes at $z>2$ is fully compatible with increased inverse Compton scattering losses at high redshifts.
\end{abstract}


%%%%%%%%%%%%%%%%%%%%%%%%%%%%%%%%%%%%%%%%%%%%%%%%%%

%%%%%%%%%%%%%%%%% BODY OF PAPER %%%%%%%%%%%%%%%%%%

\section{Introduction}
\label{sec:introduction}
High-redshift radio galaxies (HzRGs) have been found to be hosted by some of the most massive galaxies in the Universe and are thought to be progenitors of the massive ellipticals we observe today. HzRGs contain large amounts of dust and gas \citep{bes98b, arc01, car02b, reu04, deb10}, are dominated by an old stellar population \citep{bes98, roc04, sey07} and seen to have high star formation rates \citep{wil03, mcl04, mil06, vil07, sey08}. Owing to the large stellar masses of their host galaxies, radio-loud AGN across all redshifts are often found to be located in the centre of clusters and proto-clusters \citep{hil91, pen00, ven02, rot03, mil04, hat11, wyl13, ors16}. \citet{mil08} summarise the properties of radio galaxies and their environments in their review.

HzRGs are seen to be some of the most massive galaxies across all redshifts \citep[see][for example]{roc04}. As a result, near-infrared magnitudes of radio galaxies have been observed to follow a correlation with redshift, which is known as the $K-z$ relation \citep{jar01, wil03, roc04}. This is mainly because of the consistently high stellar masses observed and because HzRGs are thought to be dominated by an old stellar population, and the $K$ band traces the part of spectral energy distribution (SED) that contains light emitted from these old stars. At higher redshifts ($z>4$), the scatter in the $K-z$ relation is seen to increase as bluer parts of the rest-frame SED containing light from younger stars (which is more sensitive to the star-formation history) are observed in the $K$ band. 

Selecting HzRGs based on their radio emission alone has the advantage that no prior colour selection (one that probes the Lyman break, for example) is applied \citep{deb00}. In principle, promising high-redshift candidates selected from a radio survey with follow-up optical/IR spectroscopy could discover massive galaxies all the way into the epoch of reionisation through observations of their redshifted synchrotron spectrum. Since radio observations are not affected by dust, radio selected galaxies can also provide an unbiased sample to study the dust properties of galaxies at high redshifts. 

A successful method of identification of HzRGs from all-sky radio surveys is to first create samples of radio sources that display an ultra-steep spectral index ($\alpha$, where $S \propto \nu^\alpha$) and then follow these up at optical and infrared wavelengths \citep{tie79, rot94, cha96, blu99, deb00, ish10, afo11, sin14}. This method is effective owing to the apparent correlation that exists between the redshift of a radio source and the steepness of its low-frequency radio spectrum \citep[see][for example]{ker12}. The physical explanation for this correlation is still under debate, but it is commonly accepted that curvature of the radio SED combined with a K-correction is responsible for the apparent correlation. For a more thorough discussion on the redshift-spectral index correlation, we point the readers to \citet{mil08} where some other possible explanations are also discussed. From an observational point of view, however, radio sources with an ultra-steep spectrum (USS) generally make promising HzRG candidates. Most of the currently known radio galaxies at the highest redshifts have been selected from relatively shallow all-sky surveys. Therefore, the well-studied HzRG samples \citep{vbr99, wil99, deb00, raw01, jar01b, wil03, var10, rig11} mostly consist of luminous radio sources, and very little is currently known about the fainter and presumably more abundant radio galaxies at high redshifts. With the advent of the latest generation of radio telescopes such as the upgraded Giant Metrewave Radio Telescope (\emph{GMRT}; \citealt{gmrt}), the Low Frequency Array (\emph{LOFAR}; \citealt{lofar}), the Murchinson Widefield Array (\emph{MWA}; \citealt{mwa}) and the Square Kilometer Array (\emph{SKA}; \citealt{ska}) in the future, it is now possible to survey large areas of the sky to unprecedented depths at low radio frequencies and push the discovery of radio galaxies to lower luminosities and/or to higher redshifts, well into the epoch of reionisation (EoR; $z>6$). 

The spectra of radio galaxies across all redshifts are rich with emission lines and there have been many studies exploring the primary source of photoionisation in radio galaxies and AGN using UV and optical emission lines and their ratios. \citet{rob87} and \citet{bau92} found that photoionisation by the central AGN alone was found to sufficiently explain the UV and optical line ratios observed in low redshift radio galaxies ($z<0.2$). However, for radio galaxies at higher redshifts, photoionisation by AGN was not enough to explain the observed blue colours and bright Ly$\alpha$ emission observed in these objects \citep{vil97}. Further, there is evidence of strong interaction between the radio jet and the interstellar medium (ISM) of the host galaxy, that leads to the generation of powerful shocks that disturb the morphology, kinematics and physical conditions of the line-emitting gas \citep{oji96, mcc96, vil97}. This interaction is also supported by observations that show close alignment between the radio jet and the extended emission line regions of HzRGs \citep{mil92, jan95}. More recently, observations of the alignment of molecular gas with the radio jet in distant radio galaxies has lent further support to the jet-gas interaction scenario in radio galaxies \citep[see][for example]{emo14}. 

In this paper, we first present results from a campaign to spectroscopically follow-up the faint HzRG candidates identified by \citet{sax18a} and observe a subset of these at near-infrared wavelengths. The layout of this paper is as follows. In Section \ref{sec:sample} we present our new sample of spectroscopically confirmed radio galaxies. We first briefly describe the sample selection, observations and data reduction methods. We then present the spectroscopic and galaxy properties derived for the new radio galaxy sample, which probes fainter flux densities than previous all-sky radio surveys across all redshifts. In Section \ref{sec:extendedsample} we compile a list of known radio galaxies at $z>2$ from the literature, and use this literature sample to compare the properties of the faint high-redshift radio galaxies presented in this paper. We explore various correlations between radio properties such as luminosity, size and spectral index with emission line properties such as line luminosity and widths in both the faint and literature HzRG samples. In Section \ref{sec:discussion} we present a discussion about the possible nature of faint HzRGs and offer an explanation for the overall trends observed in the radio galaxy samples. Finally, in Section \ref{sec:conclusions} we summarise the key findings of this study.

Throughout this paper we assume a flat $\Lambda$CDM cosmology with $H_0 = 70$ km s$^{-1}$ Mpc$^{-1}$ and $\Omega_m = 0.3$.

\section{The faint HzRG sample}
\label{sec:sample}

\subsection{Sample selection}
\begin{figure}
\centering
	\includegraphics[scale=0.44]{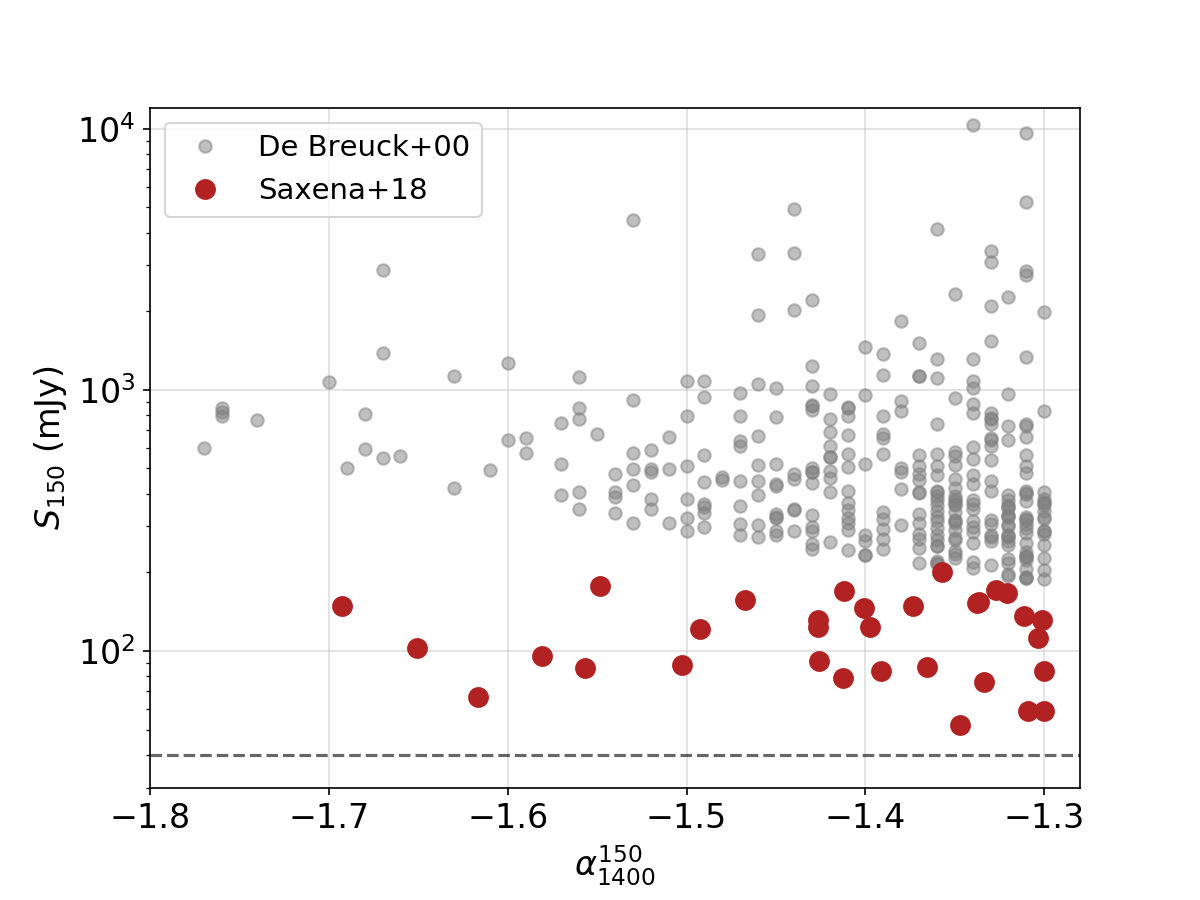}
	\caption{The unique parameter space in flux density and spectral index probed by the \citet{sax18a} sample of ultra-steep spectrum radio sources selected from TGSS ADR1. Shown for comparison are the radio sources from \citet{deb00}. Going to fainter flux densities should open up the potential for discovering radio galaxies at higher redshifts than seen before, and lead to the discovery of fainter radio galaxies at all redshifts, providing a more complete picture of the radio galaxy population.}
	\label{fig:hizesp_sample}
\end{figure}

The HzRG candidates targeted for spectroscopy are mainly selected on the basis of (i) steepness of their spectral index ($\alpha^{150}_{1400} < -1.3$) calculated by cross-matching with the VLA FIRST survey at 1.4 GHz \citep{Becker1995} and (ii) compact morphologies (largest angular size < 10 arcsec), which are expected from radio galaxies at the highest redshifts \citep{sax17}. Lower flux density limits were applied both at 150 MHz and 1.4 GHz to ensure a high enough signal to noise ratios and upper flux density limits were applied at 150 MHz to ensure that none of the sources in our final sample would have been detected in previous searches for ultra-steep spectrum radio sources \citep[for example][]{deb00}. Further, only those sources with no clear optical or infrared counterparts in the available all-sky surveys such as SDSS \citep{ala15}, PanSTARRS1 \citep{cha16}, ALLWISE \citep{wri10}, UKIDSS LAS \citep{law07} were retained in the final sample, increasing the likelihood of a high redshift nature of the radio host. The distribution of the selected sources in the radio flux density -- spectral index parameter space (taken from \citealt{sax18a}) is shown in Figure \ref{fig:hizesp_sample}.

The final sample consisted of 32 radio sources that were followed up with the Karl G. Jansky Very Large Array (VLA) in configuration A in bands L and P (Project code: 16B-309, PI: R\"{o}ttgering), with an average beam size of 1.3 arcseconds at 1.4 GHz (L-band). These results are presented in \citet{sax18a}. The high-resolution VLA images enabled sub-arcsecond isolation of the expected optical counterpart of the radio source. Blind spectroscopy at the position of the expected host of a radio source has been found to be successful at detecting emission lines and obtaining redshifts \citep{rot94, deb01}.

\subsection{Observations and data reduction}
\label{sec:obs}

\begin{table*}
\centering
\caption{Observational set-ups for spectroscopy used in this study.}
\begin{tabular}{l c c c c c c}
\hline
			&			&				&	$\lambda$ coverage	&		&	Exp. time		& Limiting flux ($1\sigma$) \\
Telescope	& Instr.		& Grism/Grating	&	($\AA$)				& Date	&	(sec)			& (\flux)			 \\
\hline \hline
WHT		&	ISIS			&	R300B/R158R	&	$3200-9200$	&	24 -- 27 Jul. 2017	&	$1800-3600$	&	$1.1 \times 10^{-18}$ 	\\
		&				&				&				&	3 -- 5 Mar. 2019	&	$3600-7200$ 	&	$2.7 \times 10^{-18}$	\\
Gemini North	&	GMOS 	& R400\_G5305 	& $5600-9400$ 	& 7 Apr. -- 24 Jun. 2017	&		$2400$		&	$1.0 \times 10^{-18}$	\\
HET		&	LRS2 IFU		&	LRS2-B/R		&	$3650-10560$	&	Dec. 2016 -- Mar. 2018	& $3600$		&	$2.0-7.0 \times 10^{-18}$	\\
\hline
\end{tabular}
\label{tab:obs}
\end{table*}

Spectra presented in this paper were obtained over a period of December 2016 -- March 2019 using three different telescopes, namely the 10m Hobby-Eberly Telescope (HET) at McDonald Observatory, Texas, the 8.2m Gemini North at Hawaii and the 4.2m William Herschel Telescope (WHT) at La Palma, Canary Islands. Due to the large distribution in right ascension of the sample, a homogeneous follow-up strategy was not practical and spectra had to be obtained over a long period of time, through multiple observing runs on numerous facilities around the world. 

We used the long-slit setup with Gemini and WHT, and the integral field unit (IFU) setup with the HET. The typical observing times were one hour on source per object. The radio morphologies for most of the targets were unresolved (at a beam size of $\sim1.3$ arcseconds and for these targets we aligned the slit at a parallactic angle of 0 degrees. For resolved radio objects, the slit was aligned along the extended radio morphology, as radio galaxy hosts are typically extended over a few arcseconds and often oriented along the radio axis \citep{rot97}. IFU observations were simply centred on the expected location of the radio host galaxy. Below we describe the observations taken using each telescope in detail. An overview of the setups used in various observing runs is given in Table \ref{tab:obs}.

\subsubsection{4.2m William Herschel Telescope}
The observations at the WHT were carried out using the Intermediate dispersion Spectrograph and Imaging System (ISIS\footnote{\url{http://www.ing.iac.es/astronomy/instruments/isis/index.html}}) instrument with a 3 arcminute slit length and a slit width of 1.5 arcseconds (Program IDs: W17AN014 and W19AN015, PI: Saxena). We used the R300B grating in the Blue arm and the R158R grating in the Red arm of the spectrograph, providing a spectral resolution of R $\sim975$ and R $\sim910$, respectively and a total spectral range of $\sim3200$ $\AA$ to $\sim9200$ $\AA$ over both arms. The first set of observations were carried out from 24 to 27 July 2017. The first two nights of the observing run were strongly affected by Saharan dust covering the island, leading to extinction of up to 2 mags in worst conditions. The conditions improved significantly for the final two nights. The average seeing over all four nights was around $0.6''$. The second set of observations were carried out from from 3 to 5 March 2019. The average seeing for the first two nights was $\sim1.0''$, and the third night was lost due to bad weather. During both runs, the same strategy for calibration was adopted, as described below.

Standard afternoon calibrations were performed and both dome and sky flats were taken before and after each observing night. A flux standard was observed at the beginning and the end of each night. Each target was observed for a total of 3600 ($1800\times2$) seconds in each arm simultaneously, except when a bright line or continuum was spotted after the first 1800 seconds. In such cases the second exposure was terminated earlier. We used CuAr + CuNe lamps for wavelength calibration, which were observed at the position of each target before the sky exposure. Since the optical magnitudes of the targets are very faint ($R > 23.5$ AB), we first acquired a bright star close to the target and then added offsets to arrive at the target coordinates. This ensures accurate positioning of the slit for very faint targets. Standard data reduction procedures that include bias subtraction, flat fielding, sky subtraction, wavelength calibration and flux calibration were performed using \textsc{python} scripts written by our team. These scripts were based on the WHT data reduction package written by Steve Crawford\footnote{can be found at: \\ \url{https://github.com/crawfordsm/wht\_reduction\_scripts}}.

\subsubsection{8.2m Gemini North}
The observations at Gemini North were carried out using the Gemini Multi-Object Spectrograph (GMOS; \citealt{gmos}) in long-slit mode with a slit width of $1.5''$ and a slit length of $5$ arcminutes (Program ID: GN-2017A-Q-8, PI: Overzier) . A total observing time of 7 hours was used to observe six targets. The observations were carried out over a period of two months, from 7 April 2017 to 24 June 2017. We used the R400\_G5305 grating with a central wavelength of 700 nm, giving a resolution of $R\sim1500$, providing coverage over the wavelength range $3600-9400$ \AA. We took three exposures of 800s each, giving 2400s of exposure time per object.

We used the Gemini IRAF package\footnote{\url{https://www.gemini.edu/sciops/data-and-results/processing-software}} for reducing the data, which performed all the basic reductions including bias subtraction, dark subtraction, flat-fielding and cosmic-ray removal. Wavelength and flux calibration was then performed on the spectra using the pipeline tools.

\subsubsection{10m Hobby-Eberly Telescope (HET)}
The observations at the HET were carried out using the Low Resolution Spectrograph 2 (LRS2; \citealt{lrs2}), which is a $6\times12$ sq. arcsec lenslet-coupled fiber integral-field unit (IFU). The wavelength coverage of the LRS2 is split into two spectrographs, the LRS2-B, which covers $3650-4670$ $\AA$ and $4540-7000$ $\AA$ at resolving powers of $R\approx2500$ and $1400$, respectively, and the LRS2-R, which covers $6430-8450$ $\AA$ and $8230-10560$ $\AA$ at resolving powers of $R\approx2500$ each. The observations were taken in spectroscopic sky conditions over four trimesters, beginning December 2016 and ending March 2018 (PI: Hill). 

We used the LRS2 quicklook reduction pipeline\footnote{\url{https://github.com/BrianaLane/LRS2_reduction}} written in \textsc{python} \citep{ind19} to perform the basic reduction of the IFU data. The reduction steps include bad pixel masking, bias subtraction, flat fielding, cosmic ray removal and wavelength calibration using arc lamps (taken with a combination of Cd, Hg, FeAr, or Kr, covering the full wavelength range) for individual frames from both blue and red arms. The reduced data products include both 2D row-stacked spectra (RSS) and 3D data cubes. The individual RSS and cubes were then median combined to form the final science data products.

\begin{figure*}
	\centering
	\includegraphics[scale=0.45]{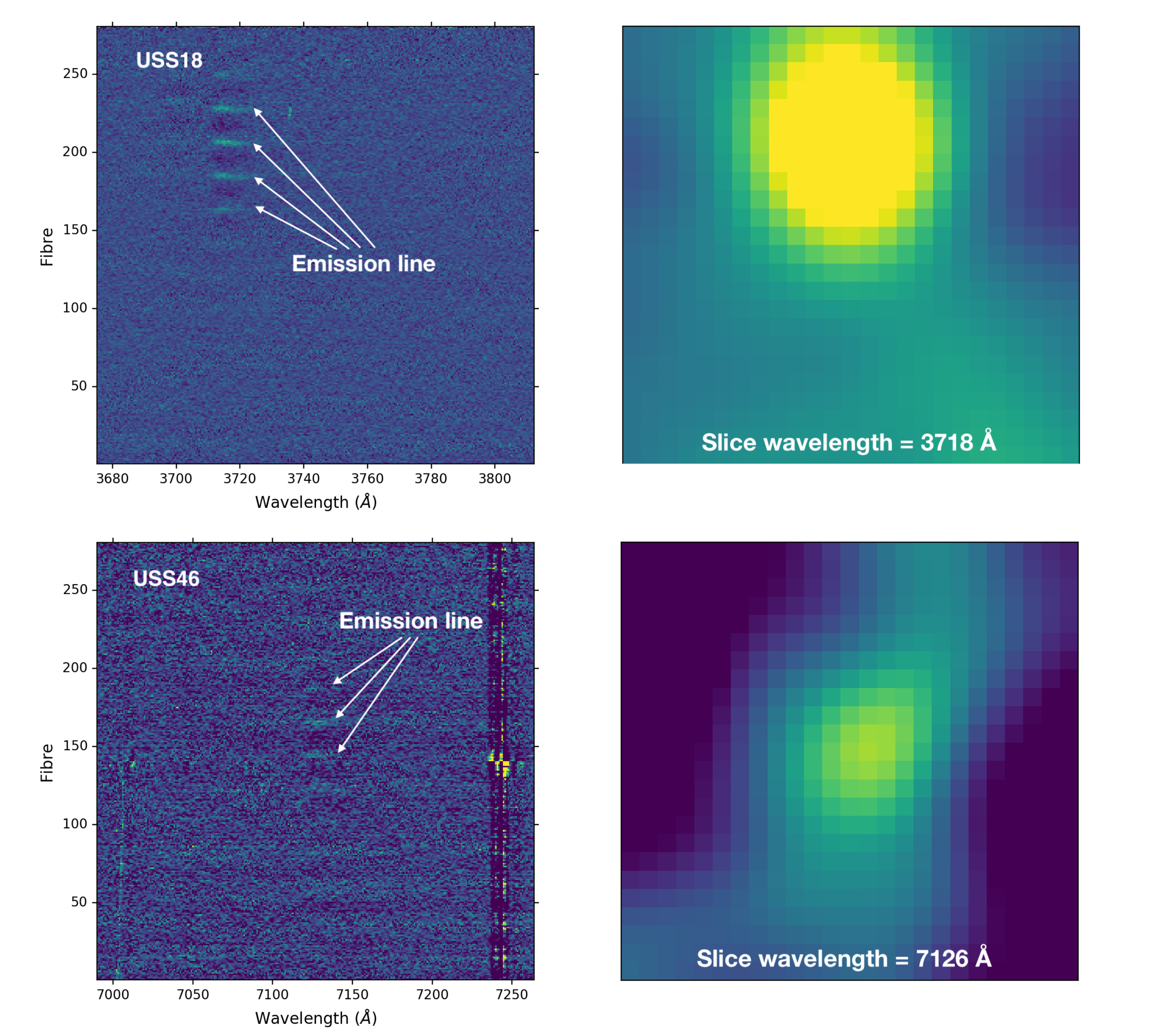}
	\caption{A visual representation of emission line identification and 1D spectrum extraction from the HET IFU data. The left hand side panels show the 2D row-stacked spectrum from the individual fibres of HET-LRS2 for a strong (top) and a weak (bottom) line emitting source, and the right hand side panels show the spatial distribution of the emission line in a slice from the 3D cube corresponding to the wavelength where the emission line flux is at its peak. The 2D spectra were used to search for emission lines, and the 3D cubes were then used to extract the 1D spectrum using a circular aperture.}
	\label{fig:het_example}
\end{figure*}

The presence of line emission in the data was first identified by visually examining the 2D spectra from each arm of the spectrograph. The 2D spectrum generated for each arm by the data reduction pipeline is essentially a row-stacked spectrum of all the IFU fibres. Presence of an emission line in an arm can be distinguished from any other spurious sources of flux by searching for signs of emission across multiple fibres. This is shown in the left hand side panels of Figure \ref{fig:het_example} for two sources. USS18 (top-left) has a strong emission line, and USS46 (bottom-left) has a weaker emission line. The distribution of emission line flux over multiple fibres helps rule out cosmic-rays or hot pixels and confirm that the emission is real. The spatial element (lenslet) size of the LRS2 IFUs is 0.6 arcseconds, so typical objects fall on several individual lenslets/fibers depending on the image quality of the observation. The spectra resulting from those fibers are separate on the CCD. Additionally, light from an unresolved emission line on each fiber falls on 14 pixels, more if the line is resolved. The combination of these factors results in a distinct pattern of separated, resolved spectra which is impossible to confuse with a CR hit (as shown in Figure \ref{fig:het_example}). Single-line detections with LRS2 are more secure than for a longslit spectrograph in this respect.

Where signs of an emission line were spotted in the 2D spectrum, we looked at the spatial distribution of the line emission in the 3D data cube generated by the pipeline for that arm of the spectrograph. This is shown in the right hand side panels of Figure \ref{fig:het_example} for both a strong and a weak emission line. A circular aperture that fully encompassed pixels with a signal-to-noise of 3 or above was then used to extract a 1D spectrum of the source. The LRS2-B and -R response functions were used to calibrate the flux of the extracted 1D spectrum and emission lines were then identified using methods that are described in the following section.

For sources with an emission line in only one arm of the spectrograph, we could not extract spectra from the other arms where no signs of emission lines were found. This is primarily because the quicklook data reduction pipeline for HET-LRS2 is unable to assign a World Coordinate System (WCS) to the individual fibres of each arm of the spectrograph, and therefore, in the absence of an emission line it is not possible to extract a spectrum by placing an aperture on a particular spatial position on the sky. A future updated pipeline will enable extraction of 1D spectra for all sources, regardless of whether or not an emission line feature has been spotted. However, since the field of view of LRS2 is relatively small ($6''\times 12''$), any detection of an emission line source is likely to be the counterpart of the radio source that has been targeted. 

Additionally, the pointing accuracy of HET is demonstrated by the detection of emission lines from sources that also have spectra from either WHT or Gemini, and by pointing tests on stars. Although we attempted to observe different targets in each of our spectroscopy runs on the three  telescopes listed above to increase the number of radio sources with redshifts, there were a few targets that were observed using multiple telescopes owing to target visibility and/or scheduling restrictions. Consistent emission line features were observed across observations using different facilities, which further helped confirm that the emission lines are real. Although specifically for the HET data cubes (with a field of view of $6'' \times 12''$, there may be an element of uncertainty associated with whether the line emitting component is actually a part of the radio galaxy. The initial radio source selection, however, takes care to eliminate any possible low-redshift interlopers.

We list the number of targets observed and the number of detections, either emission lines or continuum, recorded during all our runs in Table \ref{tab:detected}. In the next section we summarise the flux limits achieved and detection rates from each of our observing runs.

\begin{table}
\centering
\caption{Number of objects observed and number of detections during spectroscopy runs on different telescopes. There are a few objects that have duplicate observations on multiple telescopes. We also count continuum detections in these numbers, but their spectra have not been presented in this study.}

\begin{tabular}{c c c c}
\hline
Telescope (code)	&	Observed		&	Detected		&	Percentage detected	\\
\hline \hline
WHT	 (W)			&	20			&	8			&	40\%		\\
Gemini North (G)	&	6			&	5			&	83\%		\\
HET	(H)	   	 	&	15			&	8			&	54\%		\\
\hline
\end{tabular}
\label{tab:detected}
\end{table}

\subsubsection{Near-infrared imaging}
Additional near-infrared (NIR) imaging in the $J$ and $K$ bands was obtained using LUCI (formerly known as LUCIFER; \citealt{luci}) on the Large Binocular Telescope for 9 HzRG candidates in our sample. Of these, 7 had spectroscopically confirmed redshifts. Below we highlight the observations and data reduction strategy.

The observations were carried out in two separate runs, with the first on 1 February 2018 and the second on 11 May 2018 (Program ID 2017 2018 43; PI: Prandoni). The average seeing throughout the observations was $0.6-1.0$ arcseconds. The average on-source exposure times were 720 ($12\times60$s) seconds in $J$ (central wavelength of 1.247 microns) and 1200 ($20\times60$s) in $K$ (central wavelength of 2.194 microns). We used a standard jitter pattern with a maximum jitter distance of 15 arcseconds to enable effective sky subtraction in both bands.

The LUCI data reduction pipeline developed at INAF-OAR was used to perform the reductions and calibrations of raw images. These tasks include dark subtraction, bad pixel masking, cosmic ray removal, flat-fielding, and sky subtraction. Astrometric corrections for individual frames were calculated and applied and the corrected frames were then resampled and combined using a weighted co-addition method to obtain a deep image. The photometry of the co-added image was then calibrated using 2MASS and where available, UKIDSS Large Area Survey to an accuracy of within 0.1 mag. The median and standard deviation of the background in all co-added images were then calculated by placing random apertures with a diameter of 1.5 arcseconds.

One source was observed using EMIR, a near-infrared imager/spectrograph \citep{emir} on the Gran Telescopio Canarias (GTC) in queue mode (Proposal ID: 124-GTC85/17B, PI: G\'{e}nova Santos). The observation was taken in $J$ and $K$ bands, with 220 and 760 seconds of on-source exposure time, respectively. The average seeing for this observation was $1.0''$. Standard procedures for data reduction and calibrations of the raw images were performed, as mentioned previously.

\subsection{Results}
\subsubsection{Flux limits and detection rates}
The median $1\sigma$ flux limit achieved in the first observing run on WHT was $1.1 \times 10^{-18}$ \flux. We targeted 12 sources in total and were able to detect 7 (either emission lines or continuum). During the second observing run on the WHT, the conditions were inferior compared to the first run and the median flux limit achieved was $2.7 \times 10^{-18}$ \flux. We observed 9 sources in total and detected an emission line in only one source. No continuum was detected in any other source either.

The flux limits reached during observations with the Hobby-Eberly Telescope are more variable due to queue scheduling and the variable illumination during tracking. We find for our observations, which were taken across different seasons, $1\sigma$ noise levels ranging from $2.0 - 7.0 \times 10^{-18}$ \flux. On average, the blue channel is more sensitive compared to the red channel due to the lower sky background. The average flux limits reached during 6 different observations using Gemini North are $1.0 \times 10^{-18}$ \flux. The flux limits are also shown in Table \ref{tab:obs}.

\subsubsection{Redshift determination and emission line measurements}
After applying all the calibrations, the identification of emission lines in the extracted 1D spectra was attempted. The spectra of radio galaxies are dominated by one or several of the following emission lines: Ly$\alpha$ $\lambda 1216$, [O \textsc{ii}] $\lambda 3727$, [O \textsc{iii}] $\lambda5007$, H$\alpha$ $\lambda6563$ + [N \textsc{ii}] $\lambda6583$ and the initial redshift determination was based on the identification of one or more of these lines \citep{mcc93, rot97, deb01}. Relatively fainter lines such as C \textsc{iv} $\lambda1549$, He \textsc{ii} $\lambda1640$, C \textsc{iii}] $\lambda1909$, C \textsc{ii}] $\lambda2326$ and Mg \textsc{ii} $\lambda2800$ were then used in case of ambiguity arising from the identification of the stronger emission lines. For objects with a single emission line in the spectrum, we used the shape and the equivalent width of the emission line, in addition to near infrared magnitudes (where available) to identify the emission line for redshift determination. For example, an asymmetric line profile (combined with a faint near-infrared magnitude) would be a strong indicator of Ly$\alpha$ at high redshifts.  

\begin{table*}
\centering
\caption{Details of detected emission lines, radio flux densities and luminosities at 150 MHz, and radio sizes of galaxies in this paper. We mark sources with single emission lines that may have uncertain identifications with a (*). The source IDs, radio flux densities and sizes are taken from \citet{sax18a}, which also gives the RA and Dec information for each source. We mention the telescope that was used to observe each source in the Tel. column. `W' stands for William Herschel Telescope, `G' for Gemini North and `H' for Hobby-Eberly Telescope.}
	\begin{tabular}{l c c c c c c c c r}
	\hline
	Source ID		&	$z$	&	Tel.	&	Line		&	Flux 						&	Luminosity		&	FWHM	&	$S_{\textrm{150 MHz}}$ 			&		$\log L_{\textrm{150 MHz}}$ &  LAS$_{\textrm{rad}}$		\\
					&		&		&			& ($10^{-17}$ erg~s$^{-1}$ cm$^{-2}$)	& ($10^{41}$ erg s$^{-1}$) &	 (km s$^{-1}$)	&	(mJy)   &		(W Hz$^{-1}$) & (kpc)	\\
	\hline \hline
	
	USS7	& 0.53	& W	& [O \textsc{ii}] $\lambda3727$	&	4.5 $\pm$ 3.3	&	0.5 $\pm$ 0.3	&	970 $\pm$ 60	& 122 $\pm$ 24	&	26.2 $\pm$ 0.7 & 21.6 $\pm$ 0.6  \\
			&		& 	&	[O \textsc{iii}] $\lambda5007$ &	8.3 $\pm$ 2.1	&	0.9 $\pm$ 0.2	&	460 $\pm$ 30    & & & \\ 
	
	USS43	& 1.46	& W	& Mg \textsc{ii} $\lambda2800$ &	5.0 $\pm$ 3.4	& 	6.9 $\pm$ 4.7	&	640 $\pm$ 50	&   200 $\pm$ 40	&	27.6 $\pm$ 0.7 & 5.8 $\pm$ 0.8 \\
			&		&	& [O \textsc{ii}] $\lambda3727$ & 	40.0 $\pm$ 3.0 	&	55.0 $\pm$ 4.1	&	1380 $\pm$ 80	& &	 \\
			
	USS483(*)	& 1.49	& G	& [O \textsc{ii}] $\lambda3727$ & 36.6 $\pm$ 1.7  &   13.0 $\pm$ 0.6  &   1050 $\pm$ 60   & 89 $\pm$ 16   &   26.4 $\pm$ 0.7 & 18.8 $\pm$ 0.8 \\ 			
				
	USS182	& 2.02	& H	& Ly$\alpha$	            &	32.6 $\pm$ 5.0  &	96.5 $\pm$ 14.8     & 	1420 $\pm$ 60		&	146 $\pm$ 30	&	27.8 $\pm$ 0.7 & 24.2 $\pm$ 0.8 \\
	
	USS18	& 2.06	& H	& Ly$\alpha$				&	77.9 $\pm$ 4.3	&	242.0 $\pm$ 13.4    &	950 $\pm$ 60		&	131 $\pm$ 26	&	27.8 $\pm$ 0.7 & 9.5 $\pm$ 0.8  \\
	
	USS320	& 2.20	& W, H, G	& Ly$\alpha$		&	8.7 $\pm$ 3.8	&	31.7 $\pm$ 13.9     &	1300 $\pm$ 100  &   86 $\pm$ 17	&	27.9 $\pm$ 0.7 & 19.6 $\pm$ 0.8 \\
			&		& 	& Mg \textsc{ii} $\lambda2800$ &	3.2 $\pm$ 1.1	& 12.1 $\pm$ 4.1	&	320 $\pm$ 30	& & &	 \\
				
	USS268	& 2.35	& W, H	& Ly$\alpha$			&	22.8 $\pm$ 2.8  &	97.6 $\pm$ 12.0     &	750 $\pm$ 50	&   92 $\pm$ 18	&	27.8 $\pm$ 0.7 & 12.6 $\pm$ 0.8 \\
			&		&	& Mg \textsc{ii} $\lambda2800$	&	5.2 $\pm$ 3.0	&	23.2 $\pm$ 13.4 &	320 $\pm$ 30 &	& &	 \\
				
	USS98	& 2.61 	& W, G	& Ly$\alpha$			&	20.4 $\pm$ 1.6	&	112.7 $\pm$ 8.8 	&	810 $\pm$ 30	&   178 $\pm$ 36	&	28.3 $\pm$ 0.7 & 56.9 $\pm$ 0.8 \\
	USS206	& 4.01	& W		& Ly$\alpha$			&	19.5 $\pm$ 3.1	&	301.7 $\pm$ 50.0	&	1110 $\pm$ 200 &	102 $ \pm$ 20	&	28.6 $\pm$ 0.7 & 58.7 $\pm$ 0.8 \\
	
	USS172(*)	& 4.11	& H	& Ly$\alpha$				&	5.2 $\pm$ 0.7   &	85.3 $\pm$ 11.5	&	570 $\pm$ 30	&	124 $\pm$ 25	&	28.6	$\pm$ 0.7 & 65.9 $\pm$ 0.7 \\
	
	USS188	& 4.57 	& H	& Ly$\alpha$				&	6.8 $\pm$ 0.7	&	143.3 $\pm$ 14.8&	520 $\pm$ 20	&	166 $\pm$ 33	&	28.8 $\pm$ 0.7 & 26.2 $\pm$ 0.7 \\
	
	USS46	& 4.86	& H	& Ly$\alpha$				&	5.1 $\pm$ 0.8	&	124.2 $\pm$ 20.0&	500 $\pm$ 30    &	87 $\pm$ 17	&	28.7 $\pm$ 0.7 & 6.2 $\pm$ 0.6 \\
	
	USS202	& 5.72	& G	& Ly$\alpha$				&	1.6 $\pm$ 0.6	&	57.0 $\pm$ 21.4&	370 $\pm$ 30	& 170 $\pm$  34	&	29.1 $\pm$ 0.7 & 6.3 $\pm$ 0.6 \\	 
	\hline
	\end{tabular}
\label{tab:lines}
\end{table*}

Once the redshift was determined we derived the integrated flux and full width at half maximum (FWHM), by fitting a single Gaussian to the emission line. In a few cases we detect double peaked emission lines (mostly Ly$\alpha$ $\lambda1216$ and [O \textsc{ii}] $\lambda3727$), but for the purposes of this study we smoothed the spectrum in order to fit a single Gaussian to such lines. The above line identification methods yielded reliable redshifts for a total of 12 objects, which are presented in Table \ref{tab:lines}. Also shown are the flux densities and calculated radio luminosities at 150 MHz from \citet{sax18a}. The 1D spectra are shown in Figure \ref{fig:spectra} in Appendix A.

\subsubsection{Notes on individual object spectra}

\begin{enumerate}
    \item USS7, $z = 0.53$
    
    Two narrow emission lines are seen in the spectrum, which we identify as [O \textsc{iii}] $\lambda5007$ and [O \textsc{ii}] $\lambda3727$ at a redshift of $z=0.53$. Narrow [Ne \textsc{iii}] $\lambda3967$ is potentially also present, although it lies in a region of strong sky-lines. This source is slightly resolved in the radio image and the [O \textsc{iii}] and [O \textsc{ii}] line strengths suggest that this source is a faint high-excitation radio galaxy\footnote{High-excitation radio galaxies (HERGs) are a class of radio galaxies that show strong high-excitation emission lines. A characteristic feature of HERGs is their high [O \textsc{iii}] equivalent widths ($>3$ $\AA$) and high [O \textsc{iii}]/[O \textsc{ii}] ratios \citep[see][for example]{bes12}.}.

    \item USS43, $z = 1.43$
    
    A redshift of $z=1.43$ was confirmed using the [O \textsc{ii}] $\lambda 3727$ and Mg \textsc{ii} $\lambda2800$ emission lines. Also detected in the spectrum is [Ne \textsc{v}] $\lambda3426$, which has been shown to be indicative of the presence of extended emission line regions (EELRs) due to the AGN in the host galaxy, which also contribute towards powering the [O \textsc{ii}] emission in addition to star formation activity \citep{mad18}. When applying a smoothing kernel to the spectrum, we also see faint C \textsc{ii}] $\lambda2324$, C \textsc{iii}] $\lambda1909$ and C \textsc{iv} $\lambda1549$ emission. C \textsc{iii}] is blue-shifted by 30 $\AA$, which is a better indicator of the systemic redshift of the galaxy. However, since the signal-to-noise of this line is low, we quote the redshift measured using other lines instead. Our NIR imaging led to the detection of the host galaxy with $K = 21.6 \pm 0.2$ AB.
    
    \item USS483, $z=1.49$
    
    This object has an interesting spectrum with only one bright emission line visible at $9291$ $\AA$. There is faint continuum visible blueward of the emission line and the line appears to be symmetric, therefore ruling out the line identification as Ly$\alpha$ at $z\sim6.6$ where we would expect to see an asymmetric line profile and no continuum blueward of the line. 
    
    We tentatively identify this line as [O \textsc{ii}] $\lambda3727$ because of the slight hint of a double peaked morphology, giving a redshift of $z=1.49$. There is very faint emission (SNR $< 2$) visible in the spectrum that is coincident with Ne \textsc{v} at this redshift. H$\alpha$ is ruled out due to the absence of the expected [O \textsc{iii}] line in this case, and [O \textsc{iii}] is also ruled out since we would expect to see [O \textsc{ii}] $\lambda3727$ in the blue part of the spectrum. The redshift of this source remains uncertain and a deeper spectrum is required to uncover other emission lines that will help confirm the spectroscopic redshift.
    
    The long-slit was aligned along the radio axis (Figure \ref{fig:uss483_radio}, top) and the line emission in the 2D spectrum (Figure \ref{fig:uss483_radio}, bottom) is extended along the radio axis, which implies interaction between the jet and the gas surrounding the host galaxy. The line emission can be seen extending up to 15 arcseconds, pointing towards the possibility of large scale emission line halos around the central radio galaxy. 
\begin{figure}
    \centering
    \includegraphics[scale=0.3]{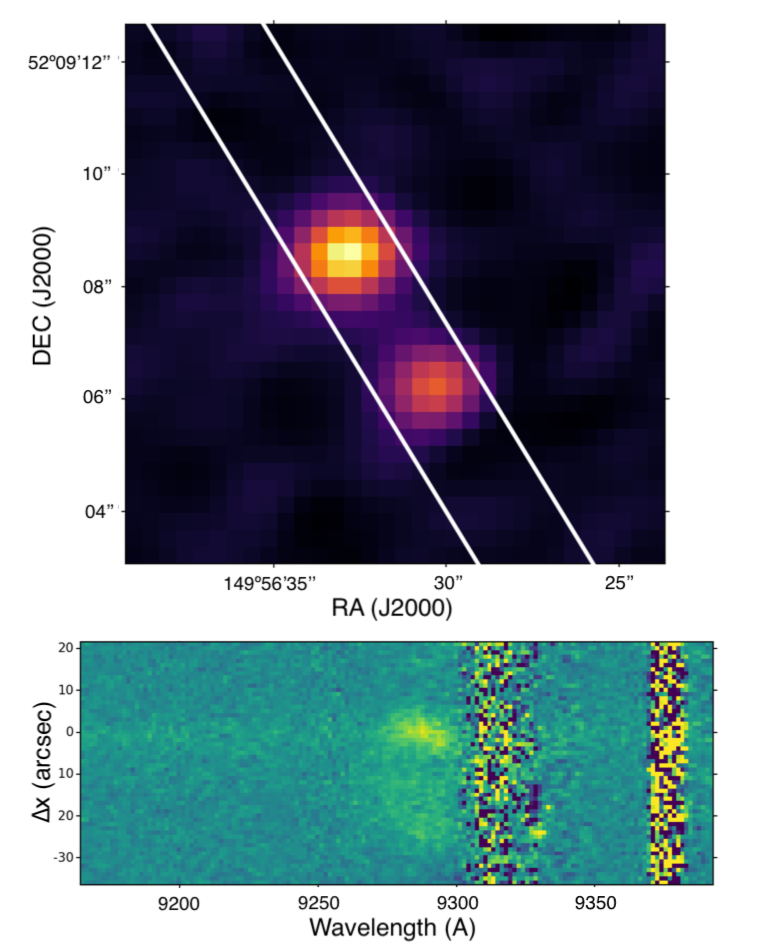}
    \caption{Top: VLA image of USS483 at 1.4 GHz, with a beam size of $\sim1.3''$. Also shown is the position of the GMOS long-slit, aligned along the radio axis. Bottom: The 2D line emission that we identify as [O \textsc{ii}] at $z=1.49$. The line emission is spatially extended along the radio axis, with emission extending all the way to $\sim15''$ away from the central host galaxy. This behaviour is a clear sign of interaction between the radio jet and the surrounding gas.}
    \label{fig:uss483_radio}
\end{figure}

    \item USS182, $z = 2.02$
    
    A strong emission line is detected in the spectrum. No other lines or continuum are detected. We identify this line as \lya at $z=2.02$ as the line is asymmetric, and lies at a wavelength that is too short for it to be \oii or any of the other bright lines typically observed in the spectrum of radio galaxies. The emission line has an integrated flux of $32.6 \pm 5.0 \times 10^{-17}$ \fluxden and a broad FWHM of 1420 $\pm$ 60 \kms. Our NIR imaging led to the detection of the host galaxy with $K=21.8 \pm 0.2$ AB.

    \item USS18, $z = 2.06$
    
    A strong emission line, which we identify as Ly$\alpha$ at $z=2.06$ is detected in the spectrum. No other lines or continuum are detected. This line is identified as \lya because of the same reasons as for USS182 -- an asymmetric line profile and too short a wavelength for \oii. The emission line has an integrated flux of $77.9 \pm 4.3 \times 1-^{-17}$ \fluxden and a FWHM of 950 $\pm$ 60 \kms. 

    \item USS320, $z = 2.20$
    
    A double-peaked emission line is observed in the spectrum, which we identify as Ly$\alpha$ at $z=2.20$. The line has a characteristic double-peaked shape with HI absorption between the two peaks, often associated with \lya from radio galaxies \citep[see][for example]{oji97}. There are also hints of very faint C \textsc{iv} and possible Mg \textsc{ii} emission (although in sky lines) in the redder parts of the spectrum at the same redshift. We smooth the double-peaked Ly$\alpha$ emission line and fit with a single Gaussian, deriving a line flux of $8.7 \pm 3.8 \times 10^{-17}$ \fluxden and a FWHM of 1300 $\pm$ 100 \kms. The host galaxy is detected with a magnitude of $K=21.7 \pm 0.3$ AB in our NIR imaging.

    \item USS268, $z = 2.35$
    
    A double peaked emission line is seen in the spectrum, which we identify as Ly$\alpha$ at $z = 2.35$. The line is identified as \lya for the same reasons as the source above -- double-peaked profile with HI absorption between the peaks. Mg \textsc{ii} is also detected in the redder part of the spectrum at the same redshift, although this is in a region affected by strong sky lines. We smooth the double peaked Ly$\alpha$ line and fit with a single Gaussian to calculate a line flux of $22.8 \pm 2.8 \times 10^{-17}$ erg~s$^{-1}$ cm $^{-2}$ and FWHM of 750 $\pm$ 50 \kms. Our NIR imaging led to the detection of the host galaxy with $J=23.6 \pm 0.1$ AB and $K=21.9 \pm 0.3$ AB.

    \item USS98, $z = 2.61$
    
    A single, strong emission line is seen in the spectrum. Owing to a lack of any other strong emission lines, we identify this line as \lya at a redshift of $z = 2.61$. The line also appears to be slightly asymmetric, strengthening the case for \lya. If the line would be \oii or \oiii, we would expect to see \oiii or H$\alpha$, respectively, in the redder parts of the spectrum. There is no clear emission line in the redder parts, and most of the expected UV lines at the \lya redshift lines fall in regions with strong sky lines. Gaussian fit to the emission line gives a line flux of $20.4 \pm 1.6 \times 10^{-17}$ \fluxden and a relatively narrow FWHM of 810 $\pm$ 30 \kms.
    
    \item USS206, $z=4.01$
    
    A single, bright and asymmetric emission line with two peaks is observed. No other emission lines or continuum are seen in the spectrum. Owing to the double peaked morphology, HI absorption signatures in the profile and a red `wing', we identify the line as Ly$\alpha$ at $z=4.01$. The measured line flux is $19.5 \pm 3.1 \times 10^{-17}$ \fluxden, with a broad FWHM of 1110 $\pm$ 200 \kms. If this line were the \oii doublet, we would expect to see \oiii in the spectrum at $\sim8200$ $\AA$. \oiii is ruled out based on the line profile as well as the absence of the expected H$\alpha$ line in the spectrum in this case.

    \item USS172, $z = 4.11$
    
    A single, narrow emission line is detected in the spectrum. No other emission lines are seen in either the blue or the red channels of the spectrograph. We tentatively identify the emission line as Ly$\alpha$ at $z=4.11$. The line is faint, with an integrated flux of $5.2 \pm 3.0 \times 10^{-17}$ \fluxden and has a narrow FWHM of 570 $\pm$ 30 \kms. We cannot definitively identify this line as \lya based on the line profile, as the line is relatively symmetric compared to the other line profiles presented in this study and the spectrum is noisier than others presented in this study. 
    
    However, this source has NIR photometry from our LBT observations, and the the host galaxy is detected with $J=22.8 \pm 0.1$ AB and $K=22.7 \pm 0.2$ AB, which are in line with expectations based on the $K-z$ relation for radio galaxies. If the line were to be \oii or \oiii, a redshift of 0.68 or 0.24 would mean that the $K$ band magnitude of the host galaxy is too faint, making it a strong outlier on the $K-z$ distribution for radio galaxies \citep[see][for example]{wil03}. The NIR colours ($J-K=0.1$) are also indicative of a high redshift nature, as a lower redshift radio galaxy would probe the rest-frame NIR where redder colours are expected \citep[see][for example]{yan15}.

    \item USS188, $z = 4.57$
    
    A single emission line is seen in the spectrum. The line is slightly asymmetric, and owing to the absence of any other strong lines in the spectrum, we identify it as Ly$\alpha$ at $z=4.57$. If this line were \oii, we would expect to see \oiii in the spectrum in the red channel of LRS2 of the HET, and in case the line were \oiii, we would expect to see H$\alpha$. The integrated line flux is $6.8 \pm 0.7 \times 10^{-17}$ \fluxden with a relatively narrow FWHM of 520 $\pm$ 20 \kms. NIR photometry of the host galaxy shows that it is undetected down to a $3\sigma$ depth of $J>23.4$ AB, which is consistent with a high-redshift nature. However, the host galaxy is detected with $K=21.6 \pm 0.2$ AB, making it relatively bright. 
    
    This bright $K$ band magnitude can be explained through expected contamination from \oii $\lambda 3727$, which would be redshifted into the $K$ band. Contribution from strong emission lines to broadband NIR magnitudes is expected from radio galaxies at certain redshifts \citep[see][for example]{mil06}. It is possible that the strong $J-K$ colour is due to a very dusty host galaxy, which would imply that detection of rest-frame UV/optical emission lines would be highly unlikely in the spectrum \citep[see][for example]{reu03b}. Since an emission line is seen in the spectrum, contamination by a strong emission line in the $K$ band is the likely explanation. 

    \item USS46, $z = 4.86$
    
    We detect a single, asymmetric emission line in the spectrum. No other emission lines or continuum are detected in the wavelength ranges covered. The asymmetric line profile, showing absorption in the blue part of the line and bumps in the red part, characteristic of \lya at high redshifts, led us to identify this line as Ly$\alpha$ at $z=4.86$. The line has an integrated flux of $5.1 \pm 0.8 \times 10^{-17}$ \fluxden and a narrow FWHM of 500 $\pm$ 30 \kms. This makes USS43 one of the highest redshift radio galaxies currently known. Again, \oii or \oiii are ruled out due to the absence of \oiii or H$\alpha$, respectively, in the redder channel of the spectrograph.

    \item USS202, $z=5.72$
    
    A single emission line, characteristic of Ly$\alpha$ at high redshift is detected in the spectrum of this object, giving a redshift of $z=5.72$. This makes it the highest redshift radio galaxy observed to date. A detailed analysis of the emission line of this particular source is presented separately in \citet{sax18b}.
    
\end{enumerate}

\subsubsection{Undetected sources}
There remain 10 USS sources that were targeted for spectroscopy and no emission lines were detected. From these, two sources have continuum detections but no emission lines, implying a possible low redshift. The rest of the sources could either lie in the redshift range where the bright optical emission lines are hard to observe from ground based telescopes ($1.5<z<2$), have passive galaxy type spectra lacking in strong emission lines, have very broad lines that were missed by our relatively shallow spectroscopy or be heavily obscured. The non-detections could also indicate that these sources are either (optically) faint, low mass radio galaxies, or lie at very high redshifts ($z>6.5$; \citealt{deb01}).

When following up samples of ultra-steep spectrum selected radio sources, there are often sources that do now show UV/optical emission lines but are detected in near-IR broadband imaging. This is usually a sign of heavy obscuration by dust, and \citet{reu03b} showed that following such sources up at sub-millimetre and millimetre wavelengths is much more efficient in detecting the host galaxy. Using such observations to constrain the SEDs, an accurate photometric redshift can be obtained along with good estimates on the physical properties of the host galaxies. Potential follow-up of the spectroscopically undetected sources at millimetre wavelengths could help identify their host galaxies and reveal their nature.

\subsection{Properties of the faint radio galaxy sample}

\subsubsection{Redshift distribution}
\begin{figure}
\centering
	\includegraphics[scale=0.44]{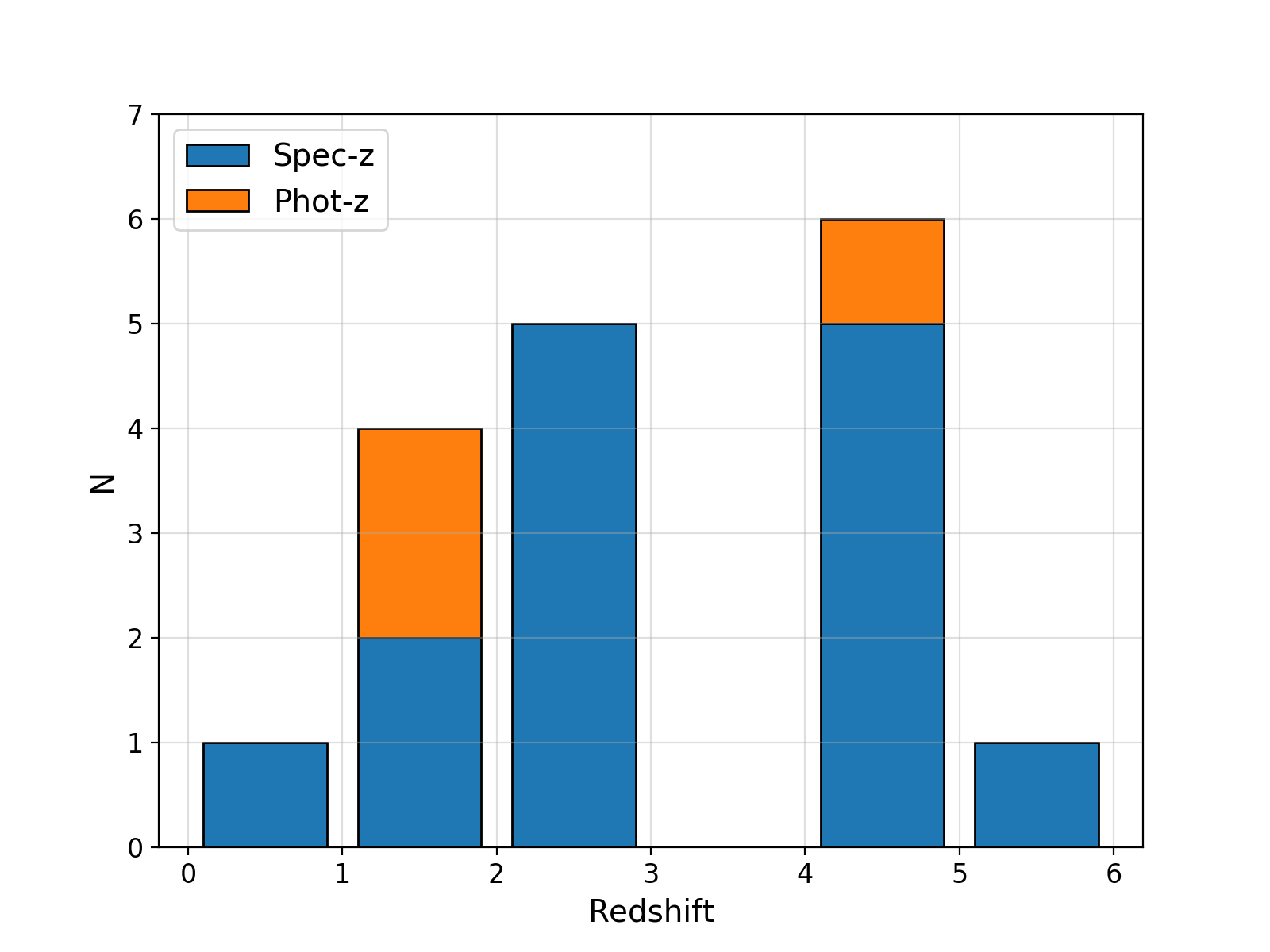}
	\caption{Stacked histogram of spectroscopic redshifts (blue) and photometric redshifts (orange) for radio sources in the \citet{sax18a} sample. The total height of the bar shows the total number of objects with either spectroscopic or photometric redshifts in each bin. The distribution shows that the USS selection is effective at picking out high-redshift radio galaxies from a large area radio survey.}
	\label{fig:redshift}
\end{figure}
In this section we look at the redshift distribution of the spectroscopically confirmed radio sources. Out of a total of 32 sources that were initially compiled by \citet{sax18a}, we spectroscopically confirmed redshifts for 13 of them. Figure \ref{fig:redshift} shows the distribution of these redshifts, along with photometric redshifts for three additional objects in the sample reported by \citet{sax18a}, which pushes the total number of sources with redshifts to 16. The total range of spectroscopic redshifts is $0.52<z<5.72$. This includes the recent discovery of the highest redshift ultra-steep spectrum radio source at $z=5.72$, which was reported separately in \citet{sax18b}. The detection of a fair number of $z>2$ radio galaxies, through USS selection combined with optical faintness and compact radio sizes, demonstrates the effectiveness of these selection criteria and holds promise for similar studies in the future with even deeper radio surveys, such as the LOFAR Two-meter Sky Survey \citep{shi17, shi19}.

\subsubsection{Velocity dispersion}
For radio sources spectroscopically confirmed in this study, through [O \textsc{ii}], [O \textsc{iii}] and Mg \textsc{ii} lines at low redshifts and the Ly$\alpha$ line at high redshifts ($z>2$), the full width at half maximum (FWHM) of the emission lines is in the range $320-1420$ km s$^{-1}$. This range is lower than what is typically observed for much brighter radio galaxies across all redshifts \citep[see][for example]{deb00} and in the following sections we compare the line widths with radio galaxies in the literature.

\subsubsection{Stellar masses}
We now estimate stellar masses for 8 (out of the 13 spectroscopically confirmed) new radio galaxies that were observed in $K$ band. To do this, we first model the expected $K$ band magnitudes for radio galaxies with different stellar masses across redshifts using the \textsc{python} package called \textsc{smpy} \citep{dun15}. \textsc{smpy} is a stellar population modelling code designed for building composite stellar populations in an easy and flexible manner and allows for synthetic photometry to be produced for single or large suites of stellar population models. To build stellar populations, we use the \citet{bc03} model with a \citet{cha03} initial mass function (IMF). We assume solar metallicity, similar to \citet{wil03}, a formation redshift of $z_f=25$ and a maximally old stellar population that has been forming stars at a constant rate. We follow the \citet{cal00} law for dust attenuation, using $A_v = 0.5$, in line with observations and predictions from \citet{mcl18} and \citet{cul18} at the redshifts and mass bins that are typical of radio galaxy hosts ($M_\star > 10^{10.2}$ $M_\odot$). The synthetic photometry produced for stellar masses in range $10.6 < \log M_\star < 12.0$ is then convolved with the $K$ band filter to calculate apparent $K$ band magnitudes over a redshift range $0<z<7$. The individual stellar mass is determined by finding the best fit stellar mass track to the $K$ band magnitude and redshift using chi-squared minimisation.

The NIR photometry for faint radio galaxies and stellar masses inferred from the models are shown in Table \ref{tab:nir}. The fainter radio galaxies have lower stellar masses on average when compared to the bright radio galaxies known in the literature. The $10^{12}$ $M_\odot$ stellar mass limit predicted for radio galaxies by \citet{roc04} stands even for our sample -- there is no radio source with an inferred stellar mass higher than $10^{12}$ $M_\odot$ in our sample. Overall, we find that fainter radio galaxies occupy the massive end of the stellar mass function at all redshifts, in line with previous observations that showed that radio galaxies are hosted by massive galaxies at every epoch \citep{roc04}. The highest redshift radio galaxy, TGSS J1530+1049 at $z=5.72$, which was part of our sample and was presented in \citet{sax18b} is one of the lowest stellar mass radio galaxy from our sample.

\begin{table}
\caption{Near infrared photometry for HzRGs and stellar mass measurements for radio galaxies in our sample.}
\begin{threeparttable}
\centering
	\begin{tabular}{l c c c r}
	\hline
	Source ID		&	$z$		&	$J$ (AB)			&	$K$	(AB)		&	$\log M_\star$ ($M_\odot$)	\\
	\hline \hline
	USS43		&	$1.46$	&	$-$				&	$21.6 \pm 0.2$	&	10.6 $\pm$ 0.4		\\
	
	USS182		&	$2.02$	&	$-$				&	$21.8 \pm 0.2$	&	11.0 $\pm$ 0.4		\\
	
	USS320		&	$2.20$	&	$-$				&	$21.7 \pm 0.3$	&	11.1 $\pm$ 0.4	\\
	
	USS268		&	$2.35$	&	$23.6 \pm 0.1$		&	$21.9 \pm 0.3$	&	11.1 $\pm$ 0.5	\\
	
	USS172		&	$4.11$	&	$22.8 \pm 0.1$		&	$22.7 \pm 0.2$	&	11.2 $\pm$ 0.5	\\
	
	USS188		&	$4.57$	&	$>23.4$			&	$21.6 \pm 0.2^{\star}$ & 11.5 $\pm$ 0.5 	\\
	
	USS202		&	$5.72$	&	$>25.3$			&	$>24.2$		&	$<10.5$		\\
	\hline
	\end{tabular}
	\begin{tablenotes}
	\footnotesize
	\item $^\star$Expected contamination from [O \textsc{ii}] $\lambda3727$ emission line in the $K$ band.
	\end{tablenotes}
	\end{threeparttable}
\label{tab:nir}
\end{table}

We note that the overall lower stellar masses of galaxies in our sample are obtained under the assumptions that go into the stellar population modelling mentioned earlier. We note that there are several key parameters that contribute towards the uncertainty on the determined stellar mass. Multi-band photometry is generally required to obtain accurate stellar masses, by constraining the SED of hosts of radio galaxies, taking into account contributions from the central AGN, stars and dust \citep[see][for example]{sey07, ove08}. Even then, assumptions regarding the IMF, formation redshift, star-formation history and the type of underlying stellar populations all contribute towards the uncertainties in stellar mass. 

Since our estimates are based on only one or two broadband photometric points, and the goal of our stellar mass measurement is to get a very simple idea based on $K$ band photometry, taking into account all the uncertainties is beyond the scope of this current analysis. However, since our models are able to account for the scatter in the $K-z$ relation for known radio galaxies without requiring extremely large or extremely small stellar masses, the inferred stellar masses for galaxies in our sample seem to be in line with expectations. Follow-up broadband photometry across optical to mid-infrared wavelengths, capturing light from the different sources that contribute to the SED, is required to put robust constraints on the stellar mass of these radio galaxies.

\section{Comparison with literature HzRG sample}
\label{sec:extendedsample}
In this section we compile a sample of known high-redshift radio galaxies ($z\ge2$) from the literature to compare the sample of newly discovered faint HzRGs against. We begin by briefly describing the parent radio samples from which the literature $z>2$ radio galaxies were selected. We particularly highlight the flux limits of each of these radio samples and note that a large majority of these radio samples have much higher flux density limits than the \citet{sax18a} sample used in this study. We point the readers to the relevant references mentioned in the sample description for detailed information about the sources.

\subsection{Literature samples}
\subsubsection{\citet{rot97}}
This sample contains spectroscopic follow-up of the Leiden ultra-steep spectrum (USS) sample \citep{rot94} and the 4C USS sample \cite{tie79}. \citet{rot97} sample contains spectra for 64 radio galaxies, with 29 of them being at $z>2$. Other Ly$\alpha$ properties from this sample are presented in \citet{oji97}. The median radio flux density of these sources is in the range $1-2$ Jy.

\subsubsection{\citet{deb01}}
This spectroscopic sample of radio galaxies follows up the USS radio sources that were first presented in \citet{deb00}. The \citet{deb01} sample contains spectra for 69 radio galaxies from the \citet{deb00} radio sample and observations of a few other previously known HzRGs, with 22 of them confirmed at $z>2$. The limiting radio flux density for various subsamples observed in this study ranges from $250-1000$ mJy.

\subsubsection{\citet{jar01b}}
This sample contains spectra for radio sources selected from the 6C* sample \citep{blu98}. The \citet{jar01b} sample has a total of 29 spectra, 12 of which lie at $z>2$. The flux density at 151 MHz of sources in this sample is in the range $0.96-2$ Jy.

\subsubsection{\citet{wil02}}
This sample contains optical spectroscopy for 49 radio sources selected from the 7C Redshift Survey, which is a spectroscopic survey of the regions covered in the 5C and 7C radio surveys. The \citet{wil02} sample contains 5 radio galaxies at $z>2$. The radio flux density probed by this sample is $S_{151}>0.5$ Jy.

\subsubsection{\citet{bor07}}
This sample contains spectra for 14 USS sources selected from the Westerbork in the Southern Hemisphere (WISH) catalogue \citep{deb02} and subsamples from the \citet{deb00} USS catalogue. Six out of 14 sources observed by \citet{bor07} lie at $z>2$. The limiting flux density of the WISH sample is $S_{352} \sim 18$ mJy.

\subsubsection{\citet{bry09}}
This sample contains redshifts for 36 galaxies, out of a total of 234 USS radio sources shortlisted from the Revised Molonglo Reference Catalogue, Sydney University Molonglo Sky Survey (MRCR-SUMMS sample) and the NRAO VLA Sky Survey (NVSS), presented in \citet{bro07}. There are 13 radio sources at $z>2$ from the \citet{bry09} sample, 11 of which have \lya line measurements. The \citet{bro07} parent radio catalogue contains sources with $S_{408} \ge 200$ mJy.

\begin{figure}
    \centering
    \includegraphics[scale=0.44]{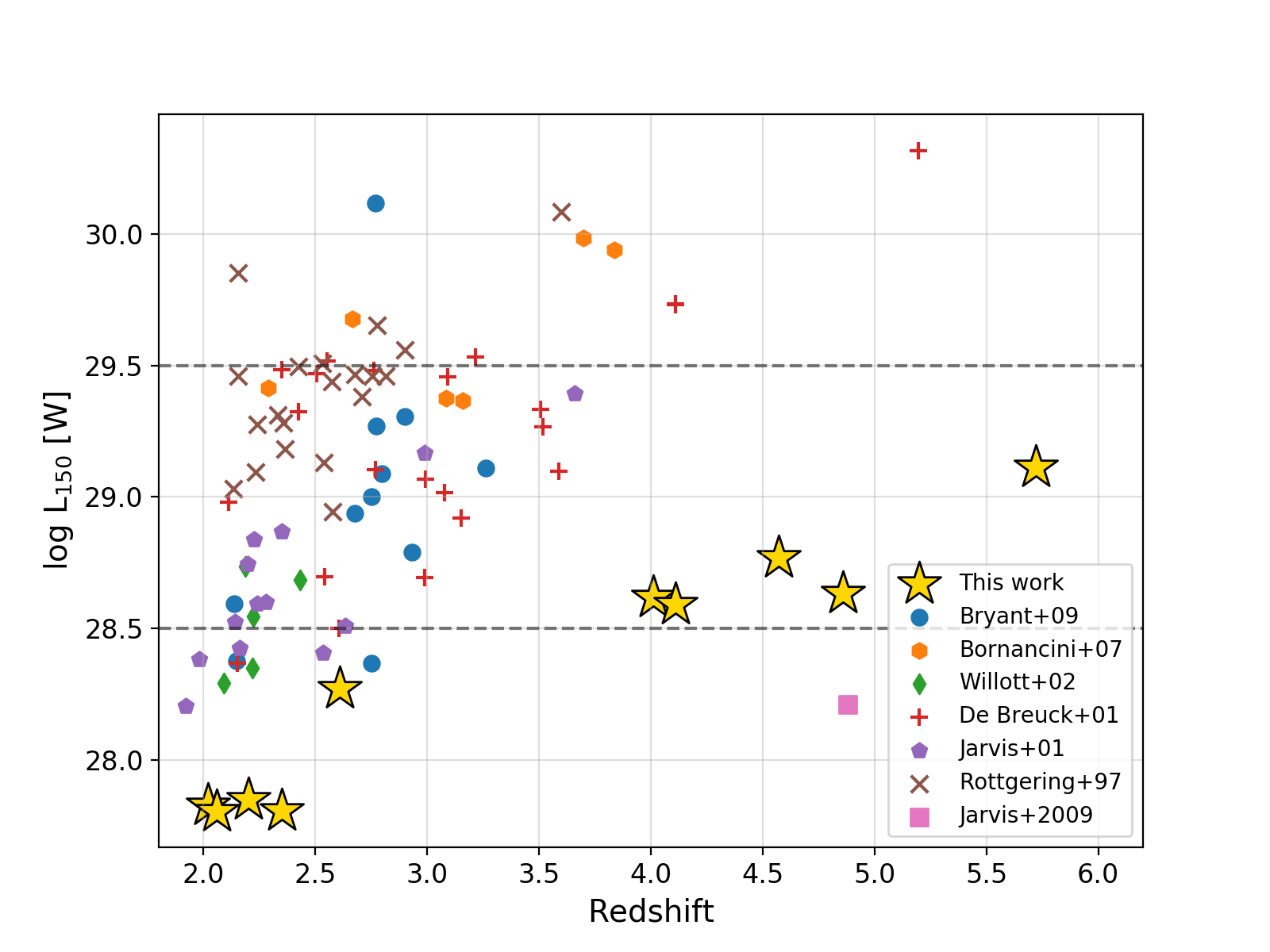}
    \caption{The distribution of radio luminosities and redshifts of the faint HzRGs (gold stars) and the literature galaxies (coloured symbols), which have been scaled to an observing frequency of 150 MHz. The faint HzRGs are almost an order of magnitude fainter at the highest redshifts, opening up a new parameter space for the discovery and study of radio galaxies in the early Universe.}
    \label{fig:lit_radiopower}
\end{figure}

\subsubsection{Radio luminosities}
\begin{figure*}
    \centering
    \includegraphics[scale=0.56]{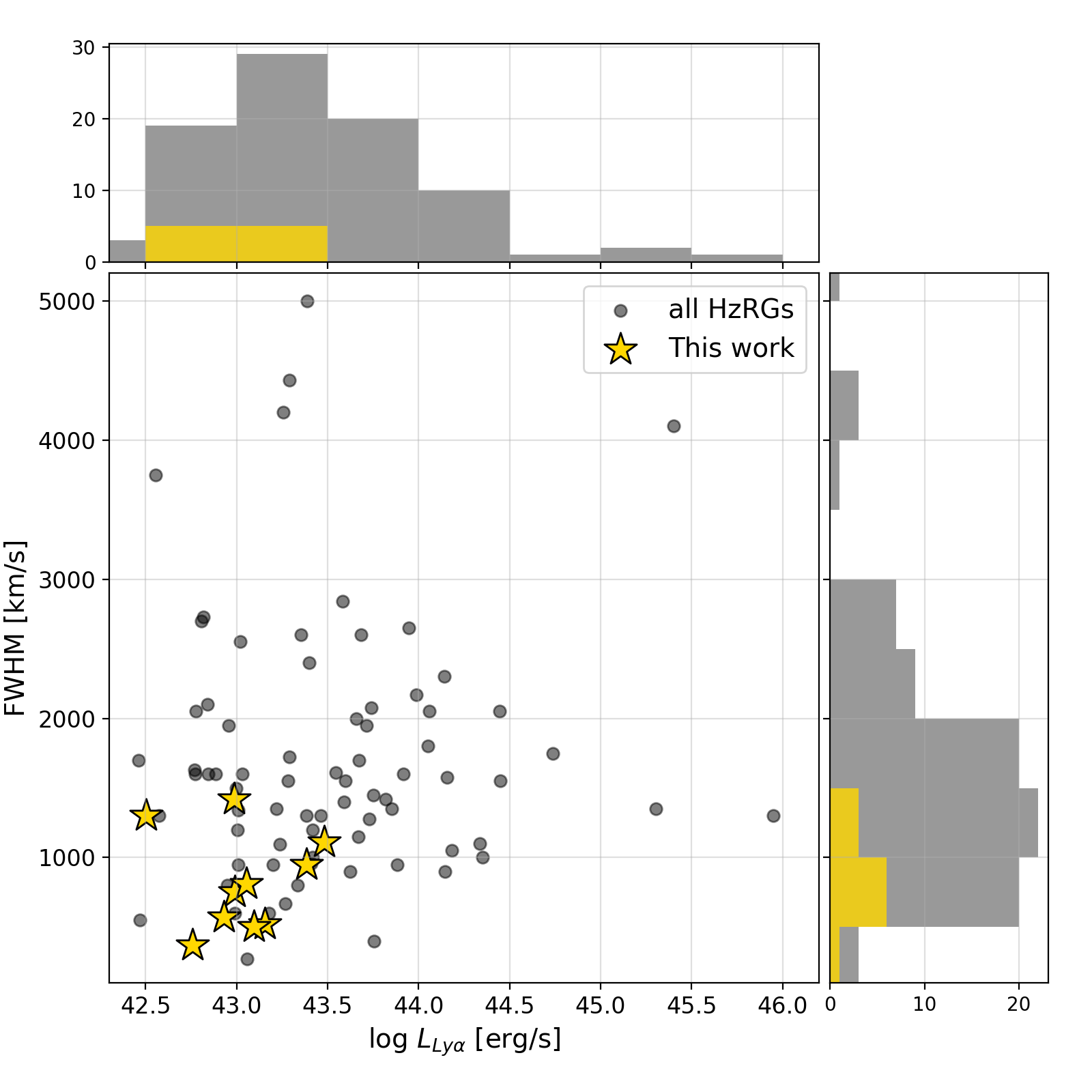}
    \caption{Distribution of Ly$\alpha$ luminosities and FWHM for faint (gold stars) and literature (black circles) radio galaxies. Also shown are the histograms of both these samples. Clearly, the faint HzRGs have lower Ly$\alpha$ luminosities and FWHM. In the text we give the median and standard deviations of Ly$\alpha$ luminosities and FWHM in the two samples. There is no correlation observed between the Ly$\alpha$ luminosities and FWHM for either of the two samples.}
    \label{fig:lya_dist}
\end{figure*}

The extended HzRG sample consists of a total of 86 radio galaxies at $z\ge2$. All these radio galaxies have Ly$\alpha$ line measurements. We calculate the radio luminosities, radio sizes and Ly$\alpha$ luminosities for the literature HzRGs using the updated cosmology used in this study. Specifically for the radio luminosities at a frequency of 150 MHz, we use the published flux density at 150 MHz wherever available and when needed, we scale their flux density to 150 MHz using the published spectral index. For the newly spectroscopically confimed radio galaxies reported in this paper, we use the radio measurements from \citet{sax18a} to calculate radio luminosities.

The distribution of radio luminosities as a function of redshift is shown in Figure \ref{fig:lit_radiopower}. Our sample of newly confirmed faint HzRGs is marked with gold stars (and in all the following sections). It is immediately clear that our sample probes radio luminosities almost and order of magnitude fainter at the highest redshifts. This includes the discovery of the highest redshift radio galaxy known till date, TGSS J1530+1049 \citep{sax18b}. We also show the faint radio galaxy detected by \citet{jar09}, which was detected in a deep field and not from an all-sky survey, as an example of the kind of sensitivites that future radio surveys would be able to achieve to probe even fainter radio galaxies at high redshifts.

\subsection{Comparison of emission line and radio properties}
In this section we will explore how the emission line and radio properties of faint HzRGs compare with those of the literature HzRG sample.

\subsubsection{Ly$\alpha$ luminosities and FWHM}
\label{sec:lyalum_fwhm}

\begin{figure}
    \centering
    \includegraphics[scale=0.44]{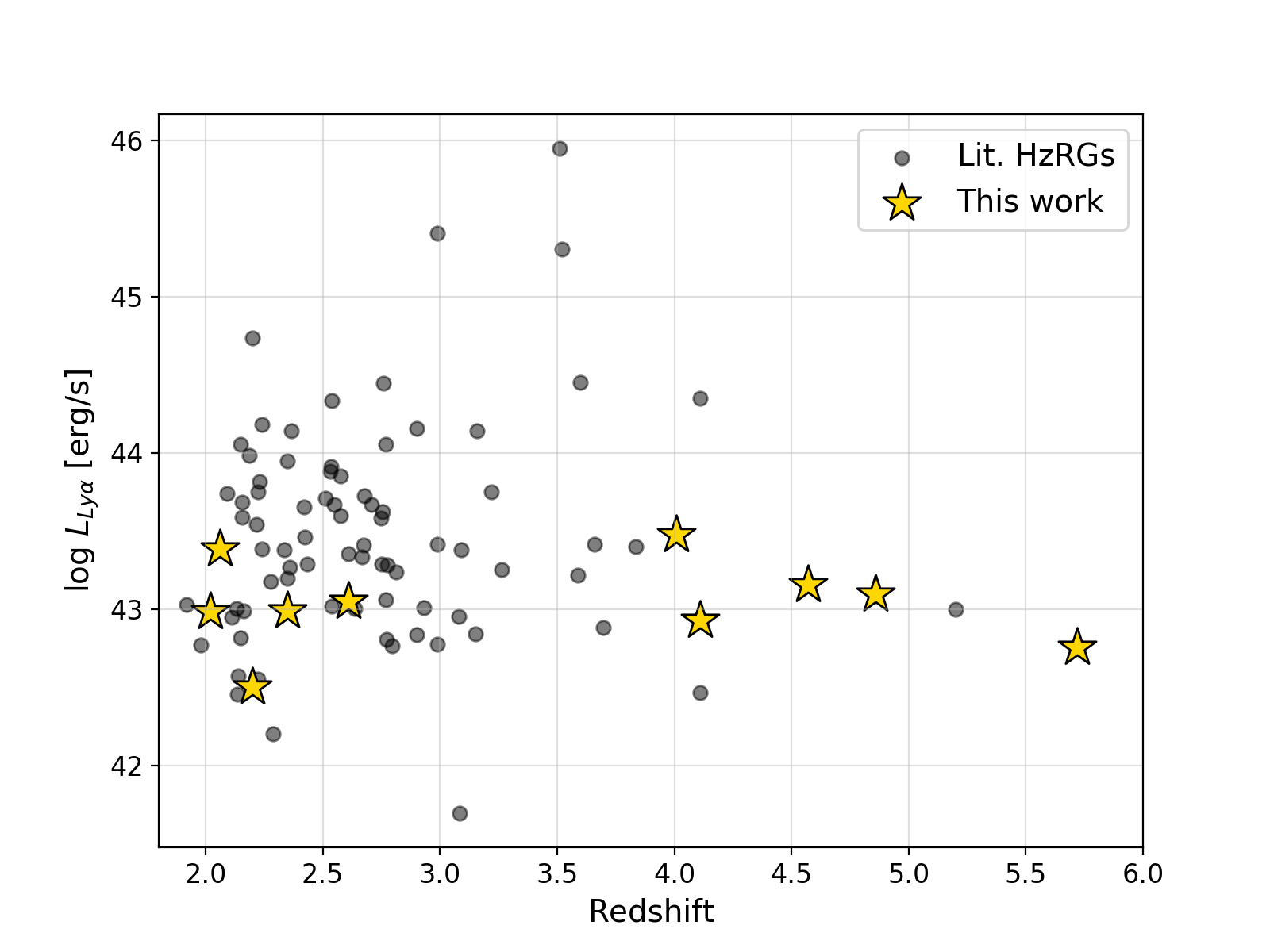}
    \caption{Ly$\alpha$ line luminosities as a function of redshift for the faint and literature HzRG samples. Across all redshifts, the faint HzRGs show lower luminosity values. We do not see any evolution in the line luminosity with redshift for either of the two HzRG samples.}
    \label{fig:lyalum_z}
\end{figure}
We begin our analysis by first showing the distribution of Ly$\alpha$ line luminosities and FWHM values for faint HzRGs (gold stars) and literature HzRGs (black circles) across all redshifts in Figure \ref{fig:lya_dist}. We note that the faint HzRGs in our sample show a systematically lower Ly$\alpha$ luminosity and FWHM compared to other HzRGs in the literature. The faint HzRG sample has a median Ly$\alpha$ luminosity of $\log L_{Ly\alpha} = 43.0$ \es, with $1\sigma$ standard deviation of $0.3$ dex. The literature HzRG sample has a median Ly$\alpha$ luminosity of $\log L_{Ly \alpha} = 43.4$ \es but a much larger $1\sigma$ standard deviation of $0.7$ dex. The difference in median FWHM values between the two samples is large, with faint HzRGs having median and standard deviation of $780 \pm 340$ ($1\sigma$) \kms, and the literature HzRGs having median and standard deviation of $1420 \pm 910$ \kms.

In Figure \ref{fig:lyalum_z} we plot the Ly$\alpha$ luminosities with redshift for both faint and literature samples. The higher dispersion in the literature sample is seen across all redshifts, with a handful of sources at $z\sim3-3.5$ showing significant deviations ($2.5-3\sigma$) from the median, such as WN J1053+5424 at $z=3.1$ exhibiting the weakest Ly$\alpha$ luminosity of the sample ($\log L_{Ly\alpha} = 41.7$ \es) and TN J0205+2242 at $z=3.5$ having the strongest luminosity ($\log L_{Ly\alpha} = 45.9$ \es). The faint HzRGs have comparable Ly$\alpha$ luminosities across redshifts. Overall, the luminosities (both faint and literature) do not seem to evolve much with redshift, with the Pearson's correlation coefficient $r=-0.02$ suggesting little to no evolution. Previously, \citet{zir09} had reported the detection of lower luminosity Ly$\alpha$ emission from a sample of $z\sim1$ radio galaxies that exhibited strong nebular emission lines, suggesting that the Ly$\alpha$ luminosities of extended halos around radio galaxies may undergo strong redshift evolution. Although our samples of both faint and literature radio galaxies do not exhibit very strong redshift evolution of Ly$\alpha$ luminosity, expanding the sample of faint radio galaxies at high redshifts may hold some clues about any such behaviour.

\begin{figure}
    \centering
    \includegraphics[scale=0.44]{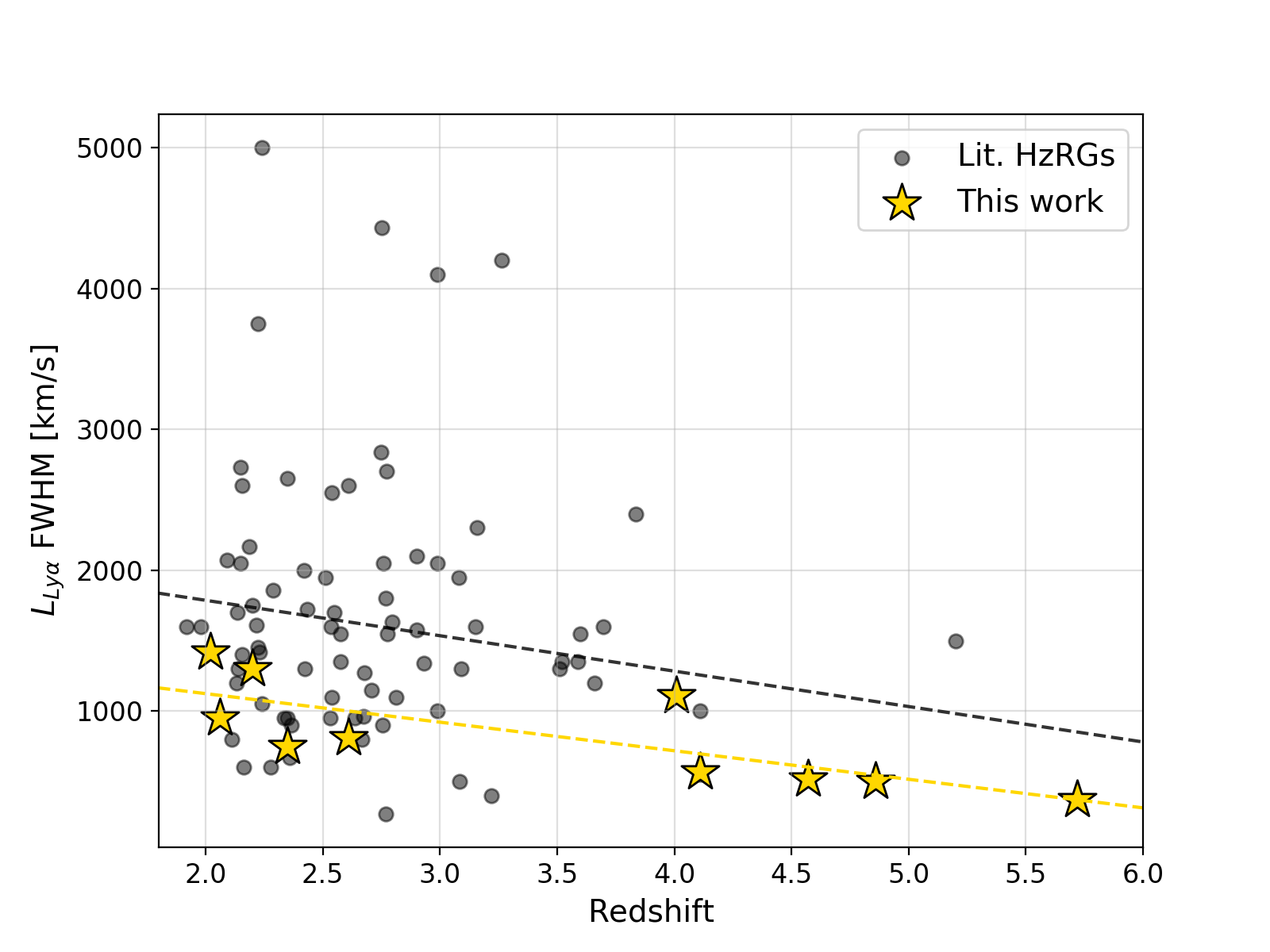}
    \caption{The Ly$\alpha$ FWHM as a function of redshift for the faint and literature HzRGs. Across all redshifts, faint HzRGs exhibit lower FWHM values when compared to the literature HzRGs. We see a decrease in FWHM with redshift in both samples, but the anticorrelation is stronger for faint HzRGs (orange dashed line) than the bright ones (black dashed line).}
    \label{fig:fwhm_redshift}
\end{figure}

Next, we plot the Ly$\alpha$ line widths (FWHM) as a function of redshift for both faint and literature HzRG samples in Figure \ref{fig:fwhm_redshift}. We find a negative correlation between FWHM and redshift for both samples and the anti-correlation is stronger in the faint sample ($r=-0.77$, compared to $r=-0.20$ for the literature sample). From the literature sample, there are 3 radio galaxies at $z\sim2-3$ that are $3\sigma$ outliers from the median value (FWHM $> 4141$ $\AA$), with the radio galaxy 1113$-$178 from \citet{rot97} showing the broadest Ly$\alpha$ profile with FWHM = 5000 km s$^{-1}$.

\subsubsection{Radio luminosity vs. Ly$\alpha$ luminosity}
\begin{figure}
    \centering
    \includegraphics[scale=0.44]{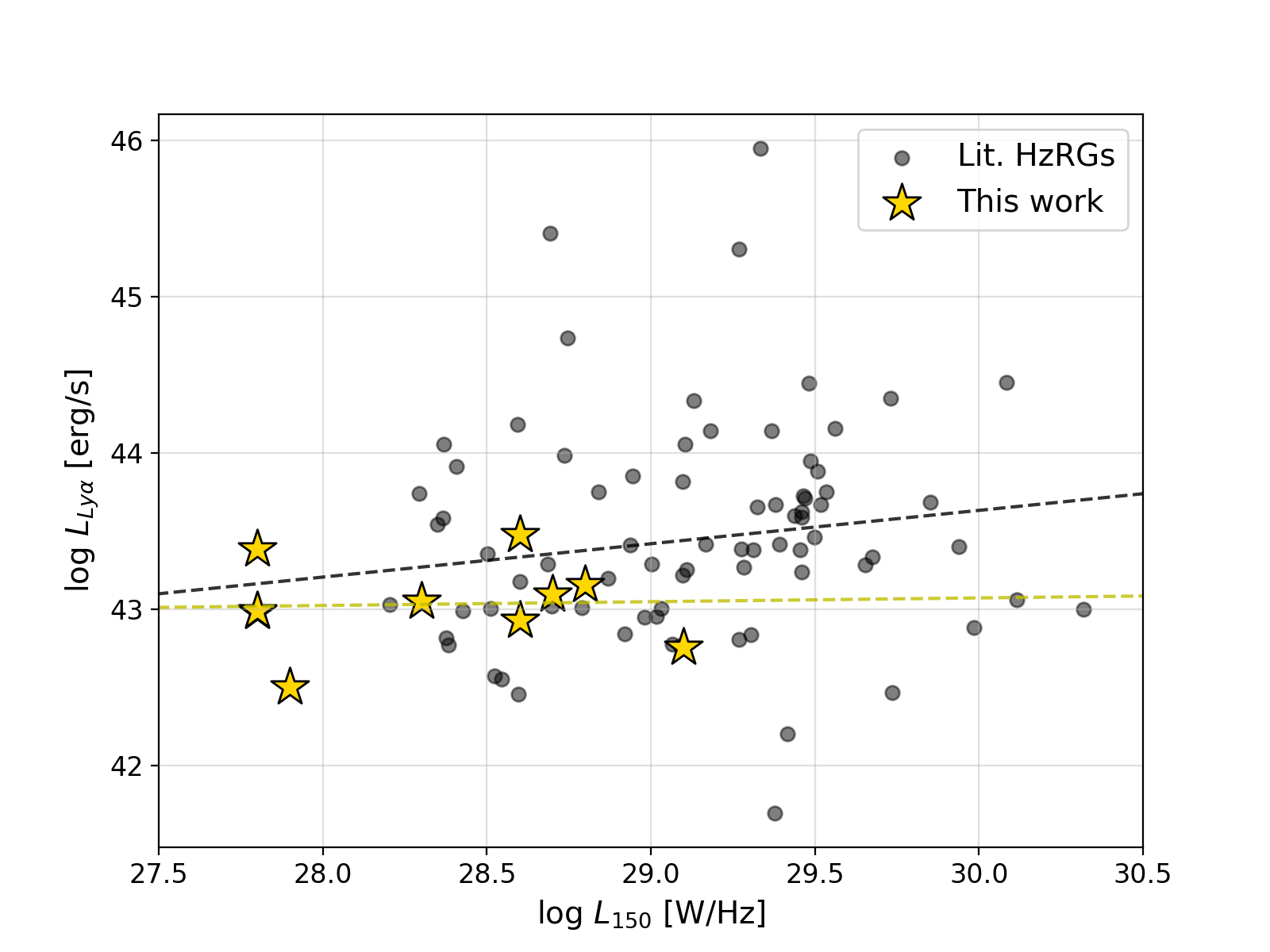}
    \caption{Ly$\alpha$ luminosities as a function of radio luminosity at 150 MHz for faint and literature HzRGs. There is a positive correlation observed between the two quantities for literature HzRGs, however it is weaker than what some previous studies have found. However, we do not see any correlation between the two for faint HzRGs.}
    \label{fig:radio_lya}
\end{figure}

We now show Ly$\alpha$ line luminosities as a function of radio luminosity at 150 MHz for the faint and literature samples in Figure \ref{fig:radio_lya}. We see a weak positive correlation in the literature sample, with $r=0.17$, which is weaker than what \citet{jar01b} observed for their sample of radio galaxies. Conversely, in the faint HzRG sample we see a very weak correlation, with $r = 0.04$, which is compatible with no correlation. Both these $r$-values, that show little to no correlation between the two quantities, indicate that the radio luminosity (across several orders of magnitude) of a radio galaxy does not have much influence on its Ly$\alpha$ luminosity. 

\subsubsection{Linear sizes}
In this section, we look at the linear size distribution of radio galaxies both samples. For faint HzRGs presented in this paper, we use the angular sizes reported by \citet{sax18a}. Linear sizes of literature HzRGs were calculated using the updated cosmology used in this study and their published angular sizes. The median size of our faint radio sources is 22 kpc and the median size of the literature HzRGs is 42 kpc. We highlight that there was angular size restriction ($<10''$) implemented by \citet{sax18b} while selecting faint HzRG candidates, which would preferentially select compact radio sources across all redshifts.

\begin{figure}
    \centering
    \includegraphics[scale=0.44]{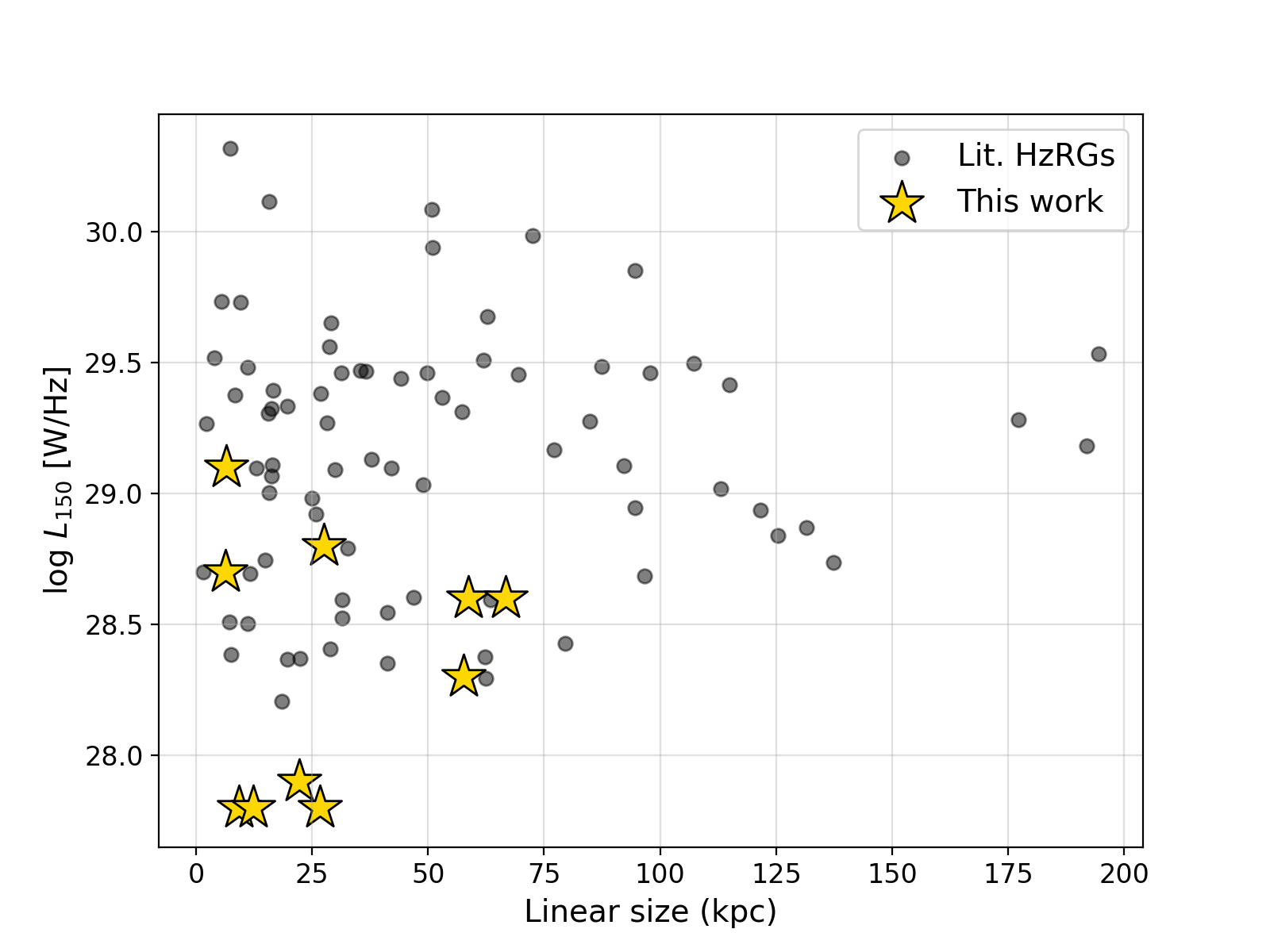}
    \caption{The distribution of radio luminosities at 150 MHz and linear sizes of sources in the faint and literature samples. the $P-D$ diagram tracks the evolution of radio sources through their lifetime. Clearly, The most luminous sources are able to grow out to larger sizes as their intrinsically more powerful jets work against the energy loss mechanisms that begin to dominate later in the radio source life. The faint HzRGs preferentially lie towards the low power, small size part of the plot, indicating that they must be young.}
    \label{fig:pd}
\end{figure}

We show the distribution of radio luminosities at 150 MHz as a function of linear size for both the faint and the literature HzRGs in Figure \ref{fig:pd}. These so-called `$P-D$' diagrams have long been used to characterise the growth and evolution of powerful radio sources across all redshifts \citep{kai97b, blu99, sax17}. The $P-D$ tracks essentially trace the evolutionary path that a radio source takes through its lifetime. For a given jet power, radio sources start out at the left side of the $P-D$ diagram. As radio sources grow, they lose energy, initially through adiabatic losses and at larger sizes due to inverse Compton losses because of interactions with the cosmic microwave background radiation \citep{kai07}. Overall, this results in the radio sources experiencing decreasing radio luminosities with increasing linear sizes, moving them towards the bottom-right side of the $P-D$ diagram. The $P-D$ tracks are influenced by magnetic field strength, redshift and the radio source environment.

As seen in Figure \ref{fig:pd}, the faint HzRGs lie in the part of the $P-D$ diagram corresponding to weaker and younger radio sources. We also note that the largest radio sources are also some of the most luminous ones and larger radio sources are expected to be older than the compact ones. Radio sources with intrinsically weaker jets are unable to grow out to large sizes as they are unable to overcome the increasing energy losses with increase in linear size, and are eventually extinguished before ever achieving large sizes \citep{sax17}. This is evident from the lack of sources in the bottom-right part of the $P-D$ diagram. We discuss the redshift evolution of linear sizes of both the faint and literature samples in Section 4.4.

\subsubsection{Radio spectral index vs. redshift}
\begin{figure}
    \centering
    \includegraphics[scale=0.44]{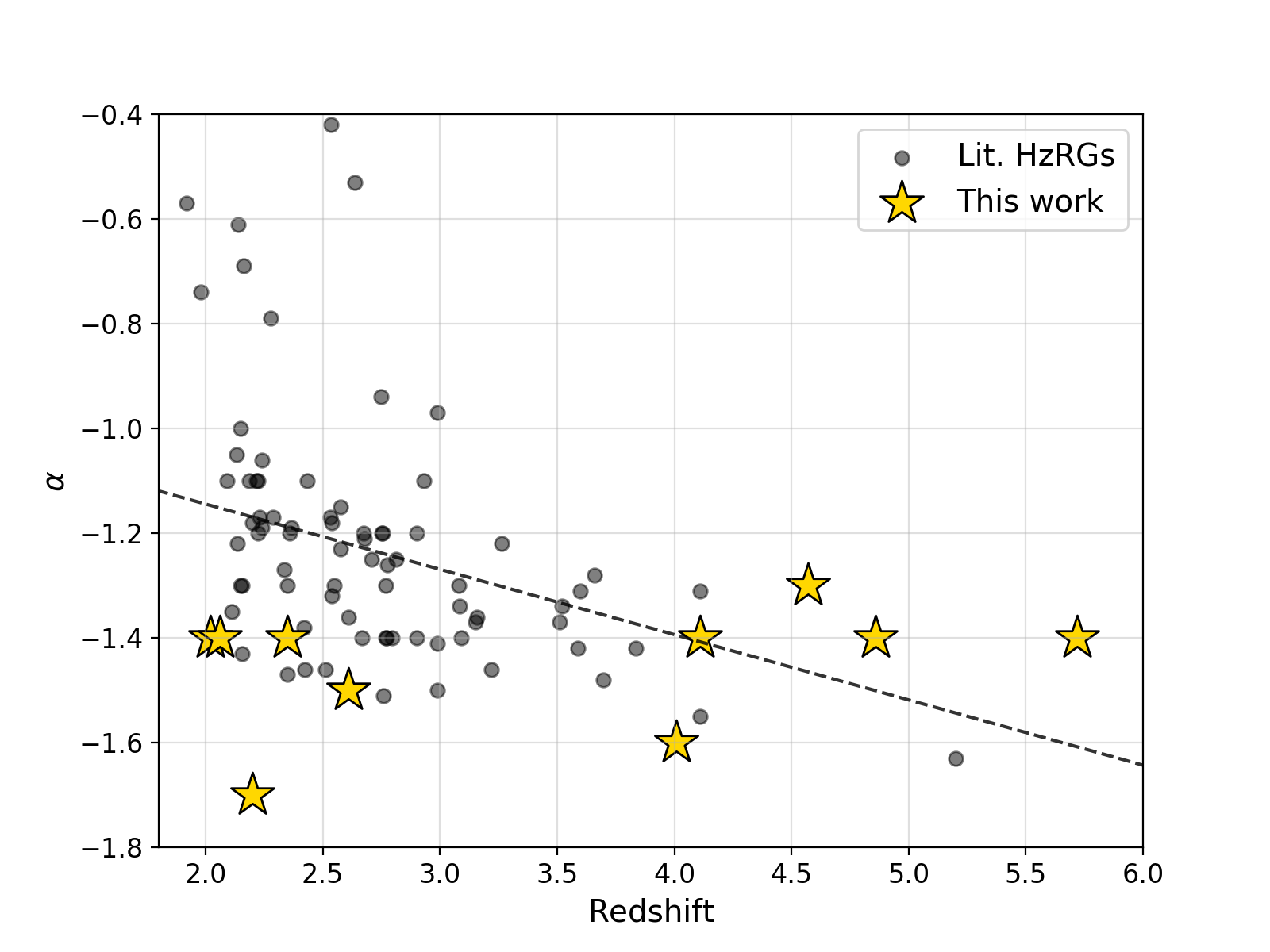}
    \caption{Low-frequency radio spectral index as a function of redshift for both faint and literature HzRGs. We observe an anticorrelation between $z$ and $\alpha$ for the literature sample, which has been used successfully to select promising HzRG candidates from large radio surveys. However, the faint HzRGs actually show little to no correlation correlation between the two quantities. Since most of the HzRGs presented in this study were initially selected because of their spectra, the observed correlations are likely to contain several biases.}
    \label{fig:zalpha}
\end{figure}

We now explore the observed correlation between the low-frequency radio spectral index ($\alpha$) and redshift ($z$) in our extended HzRG sample in Figure \ref{fig:zalpha}. The $z-\alpha$ correlation for radio galaxies has often been used to identify HzRG candidates in large area radio surveys \citep[see][for example]{rot97}.

We observe an anti-correlation between redshift and spectral index in the radio galaxies from the literature, with $r=-0.50$. However, it is important to note that most of the radio galaxies that are part of the extended sample were initially selected because of their ultra-steep spectra (USS) at low radio frequencies ($\alpha<-1$), which bias the distribution and lead to the recovered correlation. From our sample of faint galaxies, which were also selected for having ultra-steep spectra, we do not see a strong anti-correlation between spectral index and redshift. In fact, the faint radio galaxies exhibit a positive correlation between $\alpha$ and $z$, in contrast to what is seen for literature HzRGs. Further, we do not find extremely steep spectral indices at the highest redshifts, even though the best-fit from the literature would suggest so.

The origin of the observed $z-\alpha$ correlation was investigated by \citet{mor18}, where they found that a combination of inverse Compton losses at high redshifts and selection effects were able to explain the observed $z-\alpha$ correlation in radio galaxies. Inverse Compton losses affect the high frequency parts of the radio SED, which results in an overall steeper spectral index when measured at low-frequencies. Inverse Compton losses are also expected to affect the linear sizes of radio sources. However, we do not find any strong correlation between the spectral index and linear sizes of radio sources in either of the two samples studied.

\section{Discussion}
\label{sec:discussion}

\subsection{Ly$\alpha$ emission in faint HzRGs is powered by weak AGN}
\label{sec:lya}
As shown in Figure \ref{fig:lya_dist}, our newly discovered faint radio galaxies have lower Ly$\alpha$ luminosities and FWHM values across all redshifts compared to the bright HzRGs known in the literature. The dispersion in their Ly$\alpha$ luminosities is also lower. This is interesting because the distribution of Ly$\alpha$ luminosities and FWHM for brighter HzRGs shows a much larger variety (3$\sigma$ dispersion of $\log~L_{Ly\alpha} = 2.1$ \es, as opposed to $0.6$ \es for the faint radio galaxies, as shown in Section \ref{sec:lyalum_fwhm}). Historically, the large luminosities and equivalent widths of Ly$\alpha$ emission from bright HzRGs have been explained by anisotropic photoionisation from an obscured AGN in combination with jet-induced shock ionisation or contribution from hot stars \citep{cha90, bau92, mcc93, oji96, oji97, deb00b}. Additionally, physical phenomena such as launching of winds and outflows from the central AGN \citep[see][for example]{nes08}, bursts of star formation and radio AGN activity driven by recent mergers \citep[see][for example]{vil07b} and gas infall on to the host galaxy \citep[see][for example]{vil07} are sometimes required to explain the UV line ratios and large velocity dispersions at high redshifts.

The systematically lower Ly$\alpha$ luminosities in faint HzRGs indicate that the line emission must be ionised by a weaker AGN, or by modest star-burst activity. To accurately disentangle the relative contributions from AGN and star-formation, diagnostics based on UV emission line ratios are required. Several studied in the literature have showed that UV line ratios can be used to distinguish between shocks or AGN photoionisation being the dominant mechanisms \citep{vil97, all98, bes00, deb00b}. It has also been shown that UV line ratios can be used to distinguish between star-formation of AGN activity being the dominant ionising mechanism powering the lines \citep{fel16}. For the redshifts of interest presented in this study, follow-up near-infrared spectroscopy in the future offers the best way to detect rest-frame UV emission lines and test various scenarios of ionisiation. In the following sections, however, we try to set some qualitative constraints on the dominant ionising mechanisms of faint HzRGs.

\subsection{No direct evidence of strong interaction between radio jet and line-emitting gas}
Another indication of the absence of shocks is the lack of evidence for strong jet-gas interaction -- we do not see an anti-correlation between Ly$\alpha$ FWHM and linear size, which has previously been interpreted as a signature of jet-gas interaction in radio galaxies \citep{oji97, bau00, bes00}. The origin of this anti-correlation has been attributed to jets in smaller radio sources still being in the process of propagating through the relatively dense interstellar medium of their host galaxies, which leads to transfer of kinetic energy from the jets to the line emitting gas, resulting in larger line widths \citep[see][for example]{oji97}. For larger radio sources, when the jet has passed well beyond the emission line regions, the emission line widths are narrower indicating that the gas has more relaxed and settled kinematics.
\begin{figure}
    \centering
    \includegraphics[scale=0.44]{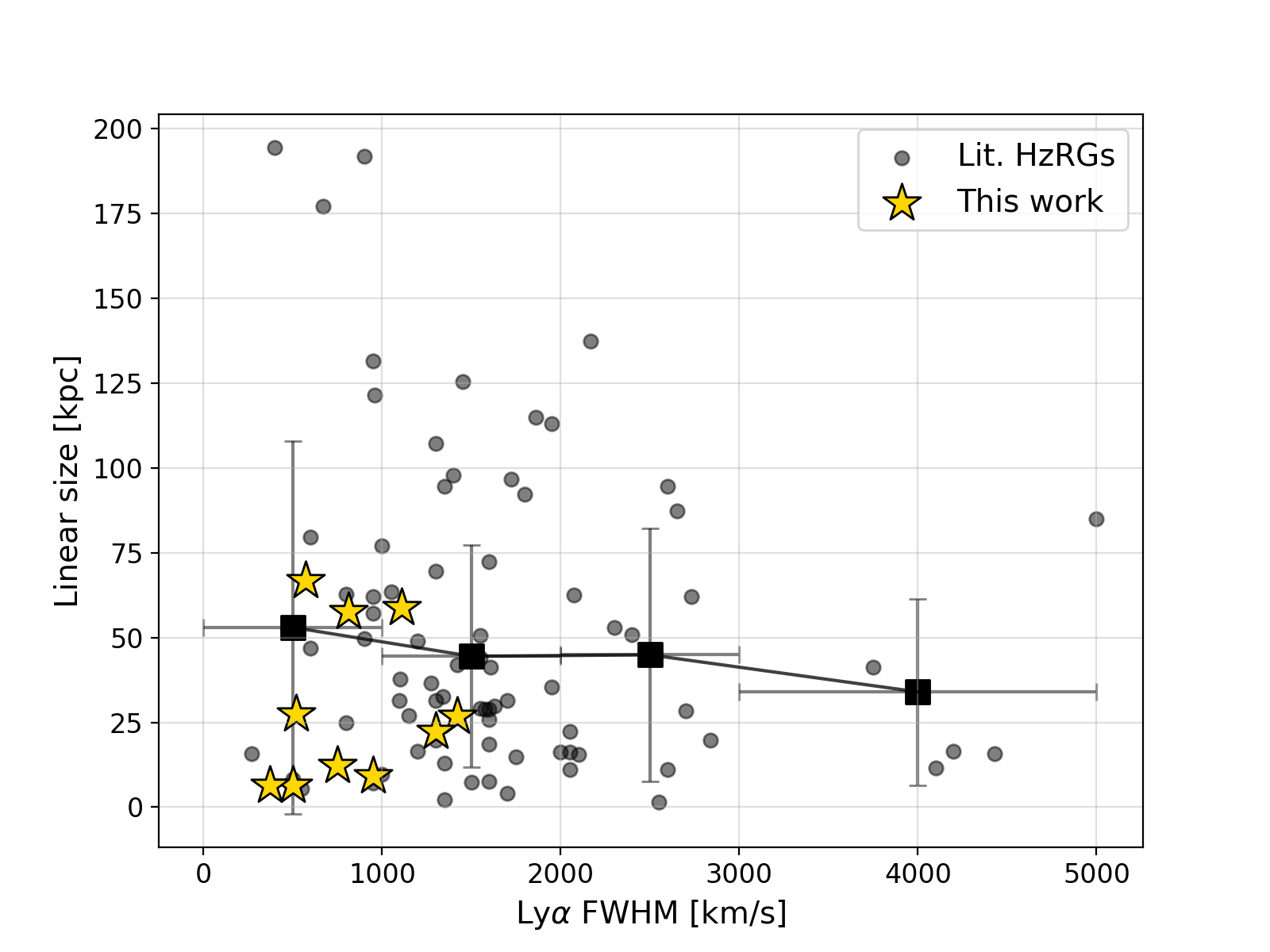}
    \caption{The linear sizes of faint HzRGs (gold stars) are not seen to anti-correlate with Ly$\alpha$ FWHM, which is indicative of absence of strong jet-gas interaction. This implies that shocks do not play a strong role in powering the Ly$\alpha$ emission in faint HzRGs. The anti-correlation in the combined sample of faint and literature HzRGs, calculated by binning in FWHM, is also very weak (black squares), although we see that the largest literature HzRGS often also exhibit narrower FWHM.}
    \label{fig:size_fwhm}
\end{figure}

In fact, when combining the faint and literature HzRG samples, it is clear that even though some of the largest radio sources in our sample show narrower Ly$\alpha$ line widths, we can not conclusively infer that narrow line widths are a consequence of less jet-gas interaction, as shown in Figure \ref{fig:size_fwhm} in which the faint HzRGs are marked with gold stars and literature HzRGs with black circles. The median linear sizes in the combined sample were calculated in bins of Ly$\alpha$ FWHM. These FWHM bins are $0-1000$, $1000-2000$, $2000-3000$ and $>3000$ km s$^{-1}$. Therefore, from our data we can not conclusively pin-point the dominant impact of jet-gas interactions in smaller radio sources. However, the added complexity of probing radio sources across a broad range of radio and Ly$\alpha$ luminosities as well as redshifts in our sample may dilute out any strong effects that may be seen in subsets of relatively similar radio galaxies. 

\subsection{Faint HzRGs have lower stellar masses}
\begin{figure}
    \centering
    \includegraphics[scale=0.44]{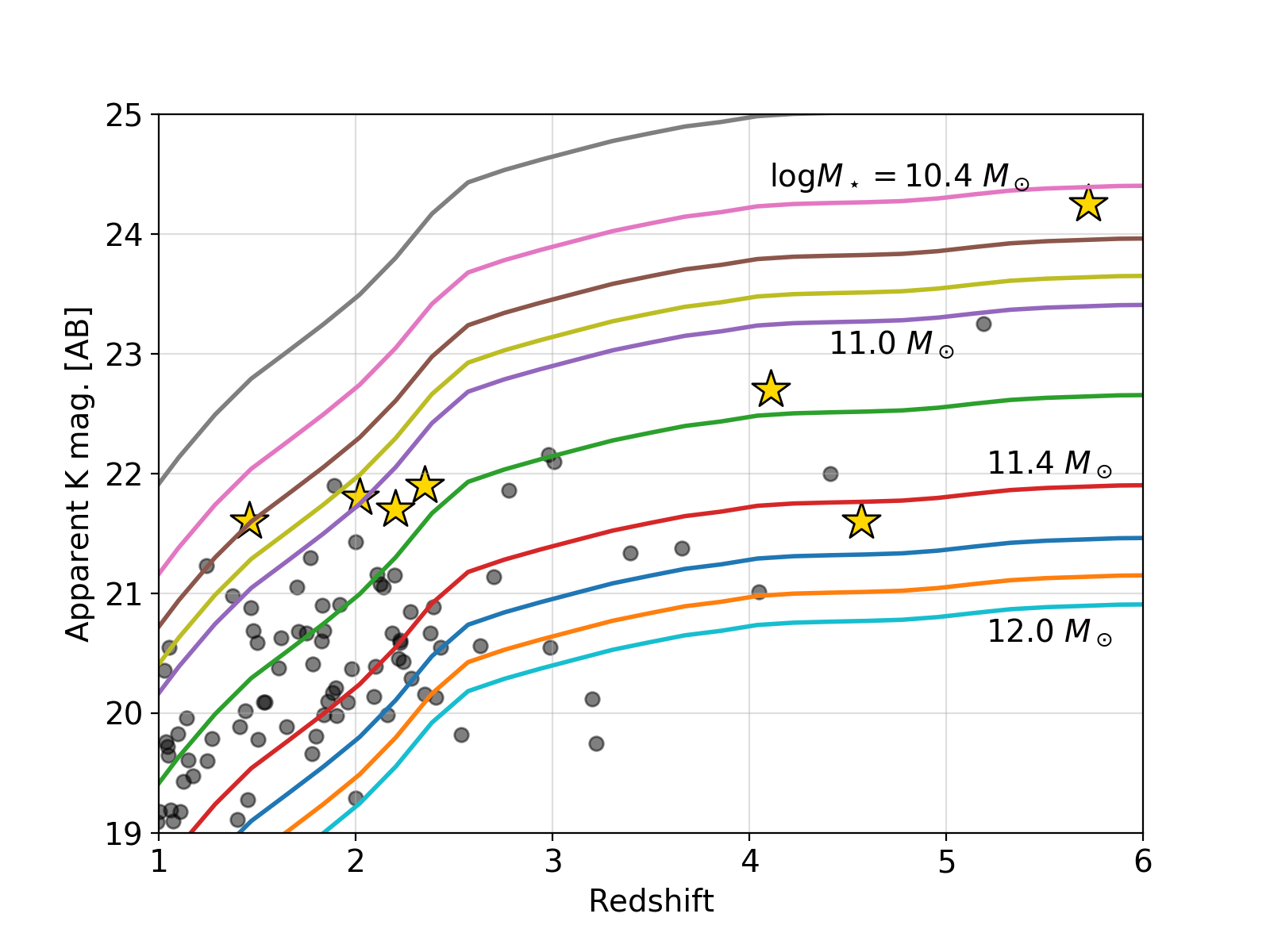}
    \caption{Apparent $K$ band magnitudes and redshifts for faint HzRGs (gold stars) and those from the literature (black dots). The solid lines indicate photometry obtained using simple stellar population modelling as described in Section 2.4.3. The solid lines show stellar mass tracks in the range $\log M_\star = 10.2-12.0$ $M_\odot$, in increments of $\log M_\star = 0.2$ $M_\odot$. Our fainter radio galaxies have lower stellar masses when compared to the previously known, more powerful radio galaxies across all redshifts. We note that TGSS J1530+1049 at $z=5.72$ has only a stellar mass limit, as it was undetected in our NIR imaging campaign \citep{sax18b}. The galaxy with the brightest $K$ mag in our sample (USS188 at $z=4.57$) has expected contamination from the [O \textsc{ii}] emission line in the broad band, which probably leads to an overestimated stellar mass using simple models.}
    \label{fig:kz_plot}
\end{figure}

As shown earlier, the apparent $K$-band magnitudes for a subset of faint radio galaxies presented in this paper were lower than what has been observed for more powerful radio sources across all redshifts. Using simple stellar population modelling, we find that the stellar masses inferred for faint HzRGs are also lower than what has been typically determined for bright HzRGs \citep[see][for example]{roc04}. This is shown in Figure \ref{fig:kz_plot}, where we show apparent magnitudes for our faint HzRGs along with those from HzRGs in the literature. 

Also shown are predictions for apparent $K$ magnitude from our stellar population synthesis modelling. Our faint HzRGs lie closer to the lower mass tracks, with the exception of USS188 with a stellar mass of $M_{\star} = 10^{11.7\pm0.7}~M_\odot$. However, contamination from the [O \textsc{ii}] $\lambda3727$ emission line is expected in the $K$ band and it is therefore likely that the stellar mass for this galaxy is being overestimated. The stellar mass estimates presented in this paper are based on only the $K$ band magnitudes and photometry in additional bands and more detailed stellar population modelling is required to accurately measure stellar masses for our sample of faint HzRGs.

\subsection{The nature of faint HzRG hosts}
\label{sec:hzrg_nature}
As shown earlier, the faint HzRGs reported in this paper overall show weaker Ly$\alpha$ luminosities and narrower line widths in comparison to the brighter HzRGs. These fainter radio galaxies are also less massive than the brighter ones. Questions naturally arise about the nature of faint radio galaxies at high redshifts. In this section we attempt to provide some qualitative constraints on this. 

To better understand the nature of fainter radio galaxies, we compare their Ly$\alpha$ line luminosities and widths with bright, `non-radio' Ly$\alpha$ emitters (LAEs) at comparable redshifts from the literature. The comparison sample is constructed mainly from LAEs that have deep spectroscopy over a wide range of luminosities, in the redshift range covered by the faint HzRGs reported in this paper. The samples we consider by no means represent the complete LAE population, but offer good statistics for comparison with our radio galaxies. 

In this analysis, we include the 21 LAEs observed by \citet{sob18} in the redshift range $z \sim 2-3$, which were initially selected from narrow-band surveys and then followed up spectroscopically. \citet{sob18} identify three broad classes of objects in their sample -- broad-line AGN, narrow-line AGN and star-forming galaxies. We also include the sample of 18 bright LAEs in the redshift range $z \sim 3-5$ observed by \citet{sai08}. This sample is divided into low equivalent width (EW$<200$ \AA) and high EW ($>200$ \AA) sources and both sub-samples contain star-forming galaxies. Finally, we include the spectroscopically confirmed brightest LAEs from the SILVERRUSH survey \citep{ouc18} presented in \citet{shi18}. These LAEs were narrow-band selected at $z=5.7$ (comparable with TGSS J1530+1049) and $z=6.6$, representing some of the brightest LAEs at $z>5.5$. Since the \lya luminosities and line widths are quite similar for the two redshifts probed by the \citet{shi18} sample, we calculate the median luminosity and line width for each redshift and use that for comparison.

\begin{figure*}
    \centering
    \includegraphics[scale=0.75]{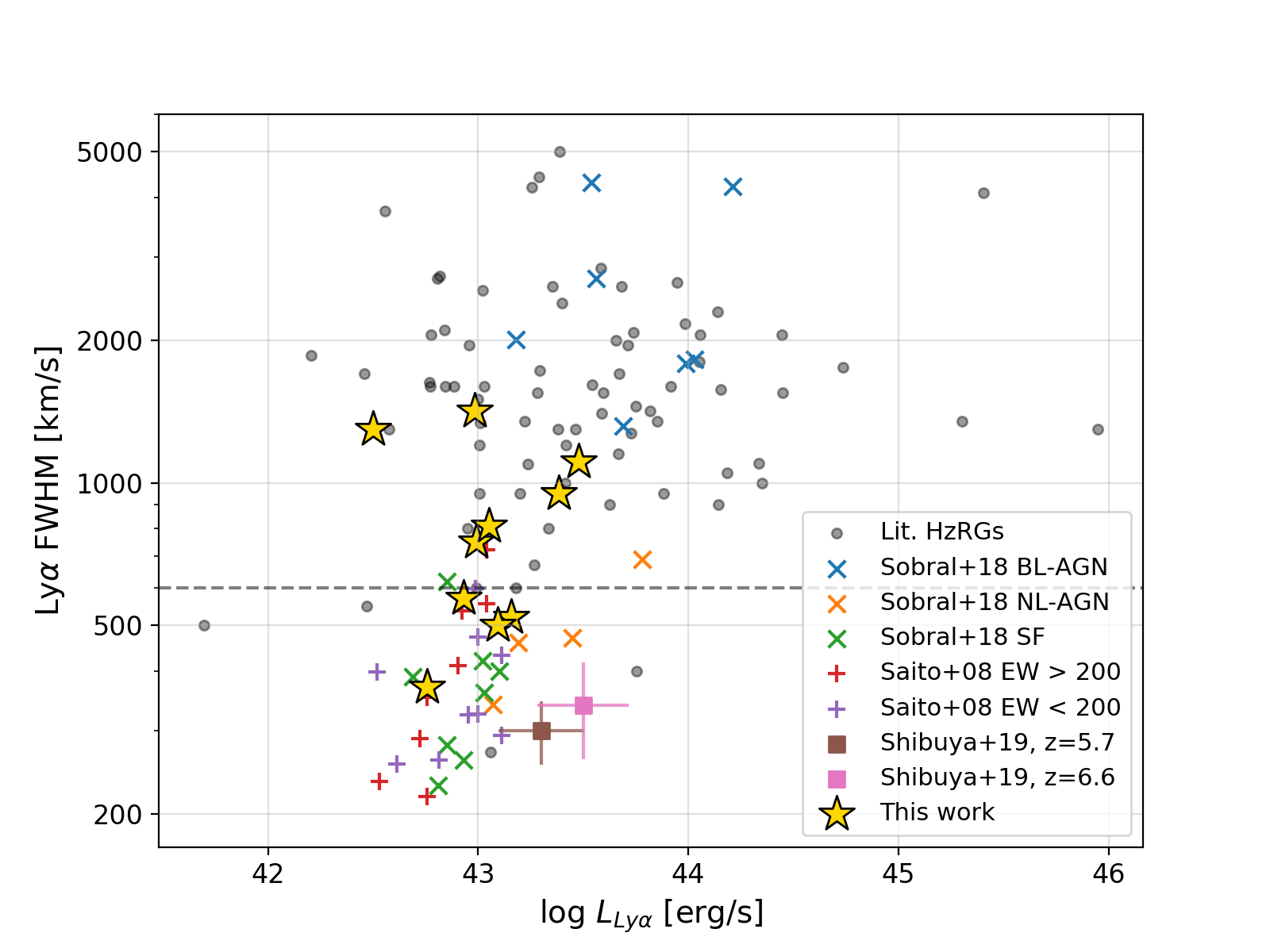}
    \caption{Ly$\alpha$ luminosity plotted against line width for faint (gold stars) and literature (black circles) HzRGs. Also shown are `non-radio' Ly$\alpha$ emitting galaxies (LAEs) from \citet{sob18} in the redshift range $z\sim2-3$ (crosses), consisting of star-forming galaxies, narrow-line AGN and broad-line AGN. We also show LAEs observed by \citet{sai08} in the redshift range $z\sim3-5$ (pluses), divided into low EW ($<200$ \AA) and high EW($> 250$ \AA) and median properties of LAEs from \citet{shi18} at $z=5.7$ and $6.6$ (squares). The black dashed line at FWHM = 600 \kms divides the faint radio galaxies at $z\sim2-4$ (lying above) and $z>4$ (lying below). We find that $z\sim2-4$ radio galaxies form a bridge between the star-forming and narrow-line AGN populations, whereas the $z>4$ radio galaxies overlap fully with the star-forming galaxy population. Our $z>4$ sources are likely powered by either moderate starburst activity or cold accretion, and must be actively building up their stellar mass.}
    \label{fig:lya_ion}
\end{figure*}

In Figure \ref{fig:lya_ion} we show the Ly$\alpha$ luminosities and FWHM for the faint HzRGs (gold stars), the literature HzRG sample (black dots), and the LAEs from \citet{sob18} (crosses), \citet{sai08} (pluses) and median values from \citet{shi18} at two redshifts (squares). Clearly, the literature HzRGs show a large scatter in both Ly$\alpha$ luminosity and FWHM values, whereas the distribution for faint HzRGs has a lower scatter. The faint HzRGs overlap with SF galaxies and narrow-line AGN from \citet{sob18} but not broad-line AGN, whereas a large number of literature radio galaxies occupy the same parameter space as broad-line AGN. Another interesting trend is the tight distribution in the parameter space of star-forming galaxies in both \citet{sob18} and \citet{sai08} samples, and 8 of our faint radio galaxies follow this trend all the way into the parameter space occupied by AGN. Four galaxies from our sample with FWHM $< 600$ \kms (below the dashed horizontal line), which all also are at $z>4$, show a complete overlap with star-forming galaxies from the LAE samples. We now try and qualitatively describe the nature of radio galaxy hosts based on what is known about the `non-radio' LAEs used for comparison. 

\subsubsection{Faint HzRGs at $z \sim 2-4$}
We first look at the fainter HzRGs presented in the redshift range $z\sim2-4$. This range consists of 6 out of the 10 sources at $z>2$ with Ly$\alpha$ measurements (lying above the dashed black line). We note that these radio galaxies occupy the regions in the parameter space in Figure \ref{fig:lya_ion} that mark the transition between the star-forming and AGN populations. For these, the \citet{sob18} galaxies offer a good sample for comparison owing to their comparable redshift range. We note that for their bright LAEs, \citet{sob18} use photoionisation modelling to show that the star-forming galaxies (green squares) have a starburst age of $\sim20$ Myrs, ionisation parameters $\log U \sim -3$ and mildly sub-solar gas metallicities. For AGN, they find an ionisation parameter $\log U \sim0.6$ and higher gas metallicities, extending up to $2Z_\odot$.

Since two or more UV emission lines are essential to accurately determine the dominant source of ionisation, and as we only detect a single Ly$\alpha$ line in a majority of our shallow spectra, we must rely on inferring ionisation properties by qualitative comparison alone. Following the measurements of \citet{sob18}, we can conclude that faint radio galaxies at $z\sim 2-4$ represent a population that is transitioning from being AGN dominated to star-formation dominated. This is in contrast to a majority of bright radio galaxies known in the literature at these redshifts that display AGN-like features \citep[see][for example]{deb01}. Further, we do not see any source that could be classified as a broad-line AGN. Therefore, faint radio galaxies in this redshift range must either be recent starbursts or narrow-line AGN, or a population transitioning from one to the other. Very little is known about faint radio galaxies at these redshifts and larger samples combined with deeper spectroscopy are essential to understand the nature of such sources.

\subsubsection{Faint HzRGs at $z > 4$}
We now look at the 4 radio galaxies (lying below the black dashed line in Figure \ref{fig:lya_ion}) that completely overlap with the star-forming LAE population. Perhaps a more direct comparison in terms of redshift for these is offered by the \citet{sai08} sources. The low EW LAEs in the \citet{sai08} sample have been identified as moderate starbursts, with star formation rates in the range $3.3-13$ $M_\odot$ yr$^{-1}$. The presence of weak AGN in these sources can not be completely ruled out. For the high EW LAEs, \citet{sai08} rule out the presence of `superwinds' launched by bursts of star formation or ionisation by a central AGN as the dominant ionising mechanism. Owing to the observed tight correlation between the line luminosities and widths, \citet{sai08} conclude that the origin of line emission in high EW LAEs is most likely due to accretion of cold gas on to the galaxy, placing them in an early phase of their evolution.

The \citet{shi18} LAEs lie at slightly higher redshifts and have much higher Ly$\alpha$ luminosities than our radio galaxies. Deep spectroscopy of these LAEs also led to non-detection of other UV lines in their spectra. \citet{shi18} rule out the presence of broad-line AGN and Population III (very low metallicity stars) as the drivers of the strong Ly$\alpha$ emission, and favour intense star-burst activity as the most likely explanation. However, the true nature of their sources remains inconclusive.

In the light of these comparison samples, we may conclude that since our $z>4$ radio galaxies coincide well with the distribution of both low and high EW LAEs from \citet{sai08} and are likely to be powered either by ordinary starbursts or by cold accretion. The \citet{sai08}, \citet{sob18} and \citet{shi18} sources lying in this parameter space do not favour AGN as the dominant ionising source. Therefore, we can say with some confidence that faint radio galaxies at $z>4$ in our sample are starburst dominated, which is different from what has been previously observed in brighter radio galaxies at these redshifts. Further, our radio galaxies show a positive correlation and a tight distribution between Ly$\alpha$ luminosity and line width, in line with the trend observed by \citet{sai08}. Therefore, we can also infer that faint HzRGs at $z>4$, if powered by cold accretion, are in an early stage of their evolution and are actively building up their stellar masses. This is consistent with the lower than average stellar masses calculated from NIR photometry for these radio galaxies.

The suggestion that faint radio galaxies at $z>4$ are actively building up their stellar mass is not entirely implausible, since radio sources at the highest redshifts are expected to be young \citep[see][for example]{blu99}. Recent VLBI radio observation of TGSS J1530+1049, a radio galaxy from our sample at $z=5.72$ \citep{sax18b}, by \citet{gab18} revealed two distinct lobes separated by only $0.4''$, confirming that the radio source is young and that the host galaxy must also be in a very early phase of its evolution.

We highlight a caveat of shallow spectroscopy though, which is possibly missing out on sources with broad emission lines. Broad lines may be present in the spectra of sources that we did not detect in our spectroscopy using relatively short exposure times, and observations of these possible broad lines may alter the inferred physical properties of faint radio galaxies that have been reported in this section.

\subsubsection{Comparison with \lya haloes seen at high redshifts}
It has been shown that several radio galaxies at $z>2$ are associated with extended Ly$\alpha$ haloes that are kinematically relaxed (\lya FWHM $\sim$ few 100s km s$^{-1}$) and often bright (\lya luminosities $\sim 10^{43-44}$; \citealt{vil03, mil06}). Deep integral-field spectroscopy (IFU), particularly with \emph{MUSE}, has indeed revealed the presence of relaxed \lya haloes around both galaxies \citep{wis16} and luminous quasars \citep{can14, arr19} at $z>3$. The existence of such haloes around the faint radio galaxies reported in this study could explain the lower than average \lya FWHM that we observe. However, we do not see any signs of extended \lya emission in the shallow spectra that we currently have. Indeed, ultra-deep sentivities are often required to probe extended haloes around high redshift objects. The \lya luminosities seen in the haloes observed in the literature also range over several orders of magnitudes and therefore, with the spectra presented in this paper, we cannot confirm the presence of such haloes around faint radio galaxies. Deep follow-up observations, either through IFU spectroscopy or narrow-band imaging at the \lya wavelengths of faint HzRGs could be the key to unveiling any such haloes in which HzRGs may be embedded.

\subsection{Inverse Compton losses dominant at high redshift}
\begin{figure}
    \centering
    \includegraphics[scale=0.44]{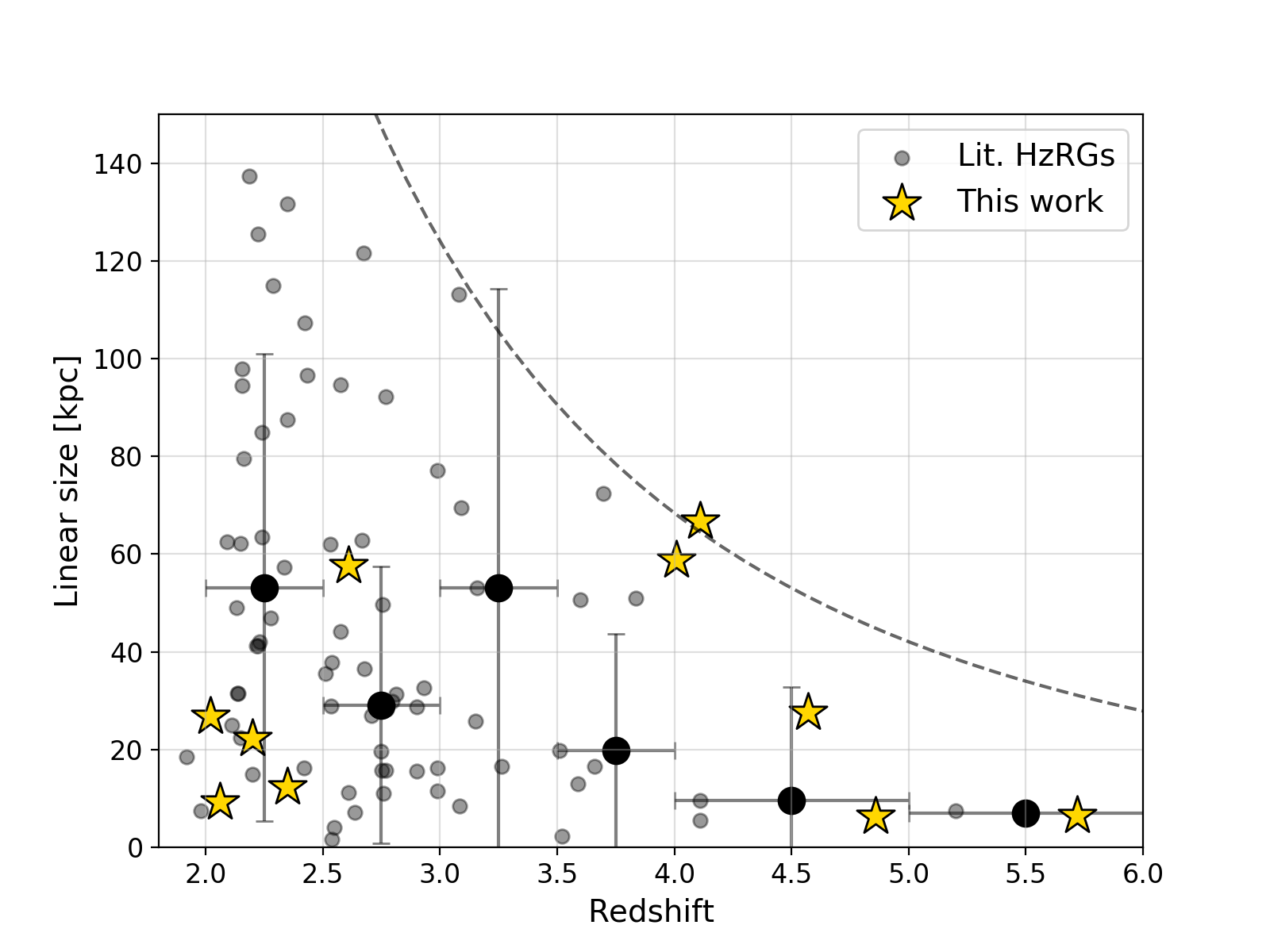}
    \caption{Linear sizes of HzRGs the faint (gold stars) and literature (black circles) samples with redshift. The median sizes, which are calculated by binning all radio sources in the combined HzRG sample in redshift bins, are shown as large black circles. Also shown are the predictions for redshift evolution of linear sizes due to inverse Compton losses, through interaction with the ambient cosmic microwave background radiation, from \citet{sax17}, that effectively provide an upper limit on achievable radio size. The median sizes of radio sources evolve strongly with redshift, and are well described by the \citet{sax17} prediction, indicating that the dominant energy loss mechanism that shapes the size distribution of radio sources at high redshifts are inverse Compton losses.}
    \label{fig:size_redshift}
\end{figure}

Although there is no strong redshift evolution seen in the sizes of the faint HzRGs reported in this paper, we do observe an evolution of radio sizes with redshift of literature radio galaxies. To gain better insights into the overall redshift evolution at $z>2$, we combine the new faint and literature samples and bin the sources in redshift bins of $z<2.5$, $2.5-3.0$, $3.0-3.5$, $3.5-4.0$, $4.0-5.0$ and $5.0-6.0$. The median linear sizes in each redshift bin as a function of redshift are shown in Figure \ref{fig:size_redshift}. The error bars on the y axis indicate the $1\sigma$ standard deviations in the redshift bins. In the highest redshift bin only two sources are present and therefore the small error bar is only indicative of $\Delta$LAS.

The strong evolution of radio sizes and the size distribution of individual sources are consistent with the prediction for the evolution of radio sizes with redshift from \citet{sax17}, where the decrease in radio size at high redshifts is primarily dominated by inverse Compton (IC) losses. This suggests that IC scattering is the dominant energy loss mechanism at high redshifts. The enhanced IC losses, due to an increasing energy density of the cosmic microwave background radiation at high redshifts, dominate over other energy loss mechanisms such as synchrotron or adiabatic losses \citep[see][for example]{sax17}. 

\section{Summary}
\label{sec:conclusions}
In this section we summarise the key findings of this paper. We have obtained spectroscopic redshifts for 13 out of the 32 sources that were selected for having an ultra-steep radio spectral index in the \citet{sax18a} sample. These radio galaxies are almost an order of magnitude fainter than those that were previously discovered from large-area radio surveys and probe a new parameter space in low radio frequency flux densities. In addition to spectroscopy we have also obtained deep near-infrared imaging for a subset of the confirmed radio galaxies. 

Next, we have compiled a sample of known $z>2$ radio galaxies in the literature that have Ly$\alpha$ measurements. The literature HzRG sample was then used to compare the properties of our faint radio galaxies across redshifts. We explored the radio and optical properties of the both the faint and literature high-redshift radio galaxies The main conclusions of this study are as follows:

\begin{enumerate}
	\item The newly confirmed fainter radio galaxies lie in the redshift range $0.52\le z \le 5.72$, including the discovery of the highest redshift radio galaxy discovered to date, TGSS J1530+1049, which was first reported in \citet{sax18b}. The Ly$\alpha$ emission line profiles show a wide variety of FWHM values, ranging from $320 - 1240$ km s$^{-1}$. The Ly$\alpha$ luminosities and FWHM are systematically lower than what has previously been observed for brighter radio galaxies at these redshifts. Due to fainter line luminosities, narrower line widths and no evidence of ionisation by shocks in the faint HzRGs, we conclude that Ly$\alpha$ emission in these galaxies must be powered by weak AGN.
	
	\item We use $K$ band photometry for a subset of spectroscopically confirmed faint HzRGs presented in this paper to estimate stellar masses. Simple stellar population synthesis models are used for this. We find that the stellar masses to lie in the range $10.7 < \log M_\star <11.7$ $M_\odot$, which are lower than what is seen for brighter HzRGs in the literature. Therefore, fainter HzRGs are hosted by galaxies with lower stellar masses, which are in a phase of building up their stellar mass rapidly.
	
	\item We do not find any strong evidence for the presence of shocks in our sample of faint HzRGs. Therefore, we conclude that the dominant sources of ionisation in faint HzRGs must either be weak AGN or star-formation. We compare the Ly$\alpha$ line luminosity and width with strong, `non-radio' Ly$\alpha$ emitting galaxie and find that faint HzRGs in the redshift range $z\sim2-4$ form a bridge between the star-forming and narrow-line AGN populations. They do not show broad-line AGN characteristics. At $z>4$, faint HzRGs completely overlap with the star-forming galaxy population, from which we infer that they could be powered by moderate star-burst activity. It is also possible that $z>4$ radio galaxies are powered by cold accretion, which in combination with their low stellar masses shows that they must be in the process of assembling their stellar mass.
	
	\item The linear sizes of faint HzRGs do not evolve much with redshift, but the sizes of the literature HzRGs exhibit a decrease with increasing redshift. We find that the redshift evolution of sizes of sources in a combined sample of faint and literature HzRGs matches well with predictions of size evolution due to increased inverse Compton losses at high redshifts. Therefore, the dominant energy loss mechanism in radio galaxies at high redshifts is the energy loss due to interactions with the cosmic microwave background radiation, and not synchrotron or adiabatic losses.

\end{enumerate}

There remain many radio sources from the original \citet{sax18a} sample that still do not have a spectroscopic redshift and are high-priority targets for spectroscopy. Our new sample of HzRGs, that probes fainter flux densities over a large area on the sky, has offered glimpses into the nature of relatively less powerful radio galaxies at high redshifts. Ongoing large area surveys that are even more sensitive at low radio frequencies, such as the LOFAR Two-Meter Sky Survey \citep{shi17, shi19} will uncover many more faint radio galaxies at high-redshifts, furthering our understanding of how radio galaxies are fuelled and their impact on their surroundings.

\section*{Acknowledgements}
We thank the referee for comments and suggestions that helped improve the quality of this work. AS wishes to thank Ricardo T. G\'{e}nova Santos for fruitful discussions. AS, HJR and KJD gratefully acknowledge support from the European Research Council under the European Unions Seventh Framework Programme (FP/2007-2013)/ERC Advanced Grant NEWCLUSTERS-321271. RAO and MM received support from CNPq (400738/2014-7, 309456/2016-9) and FAPERJ (202.876/2015). IP acknowledges funding from the INAF PRIN-SKA 2017 project 1.05.01.88.04 (FORECaST). PNB is grateful for support from STFC via grant ST/M001229/1.

This paper contains data from several different facilities. The William Herschel Telescope is operated on the island of La Palma by the Isaac Newton Group of Telescopes in the Spanish Observatorio del Roque de los Muchachos of the Instituto de Astrof\'{i}sica de Canarias. The Hobby-Eberly Telescope (HET) is a joint project of the University of Texas at Austin, the Pennsylvania State University, Stanford University, Ludwig-Maximilians-Universit{\"at} M{\"u}nchen, and Georg-August-Universit{\"a}t G{\"o}ttingen. The HET is named in honor of its principal benefactors, William P. Hobby and Robert E. Eberly. The Gemini Observatory is operated by the Association of Universities for Research in Astronomy, Inc., under a cooperative agreement with the NSF on behalf of the Gemini partnership: the National Science Foundation (United States), the National Research Council (Canada), CONICYT (Chile), Ministerio de Ciencia, Tecnolog\'{i}a e Innovaci\'{o}n Productiva (Argentina), and Minist\'{e}rio da Ci\^{e}ncia, Tecnologia e Inova\c{c}\~{a}o (Brazil). The Large Binocular Telescope (LBT), an international collaboration among institutions in the United States, Italy and Germany. LBT Corporation partners are: The University of Arizona on behalf of the Arizona university system; Istituto Nazionale di Astrofisica, Italy; LBT Beteiligungsgesellschaft, Germany, representing the Max-Planck Society, the Astrophysical Institute Potsdam, and Heidelberg University; The Ohio State University, and The Research Corporation, on behalf of The University of Notre Dame, University of Minnesota, and University of Virginia.

This work has made extensive use of \textsc{ipython} \citep{per07}, \textsc{astropy} \citep{ast13}, \textsc{aplpy} \citep{apl}, \textsc{matplotlib} \citep{plt}, \textsc{mpdaf} \citep{mpdaf} and \textsc{topcat} \citep{top05}. This work would not have been possible without the countless hours put in by members of the open-source developing community all around the world. 

%%%%%%%%%%%%%%%%%%%%%%%%%%%%%%%%%%%%%%%%%%%%%%%%%%



\bibliographystyle{mnras}
\bibliography{spec_hzrg} % if your bibtex file is called example.bib

%%%%%%%%%%%%%%%%%%%%%%%%%%%%%%%%%%%%%%%%%%%%%%%%%%

%%%%%%%%%%%%%%%%% APPENDICES %%%%%%%%%%%%%%%%%%%%%

%%%%%%%%%%%%%%%%%%%%%%%%%%%%%%%%%%%%%%%%%%%%%%%%%%

\appendix

\section{Long-slit spectra}
Here we show the 1D spectra (observed frame) of all confirmed radio sources observed using three different facilities. The spectra are displayed with increasing redshift, and contain information about the source name, the redshift and the telescope used. Regions of strong skylines are marked by shaded boxes. We show a zoom-in of the \lya emission line for radio galaxies with $z>2$. These insets are placed on regions strongly affected by sky-lines, so that the good parts of the spectrum are always visible.

The spectra have been normalised by the strength of the strongest emission line visible in the spectrum (\lya when available). Depending on the instrument used to obtain the spectrum, the wavelength range visible varies. The ISIS instrument on WHT has two arms (Blue and Red), providing the largest wavelength range. The GMOS spectra only have the red arm. The HET spectra are split into 2 arms (Blue and Red), with each arm having 2 channels. Because LRS2 on HET is an integral-field unit spectrograph, we could only extract the 1D spectra from those channels that had any visible continuum or emission lines. Therefore, we only show the spectrum of the channel where any emission was visible.
\begin{figure*}
\centering
\includegraphics[width=0.95\textwidth]{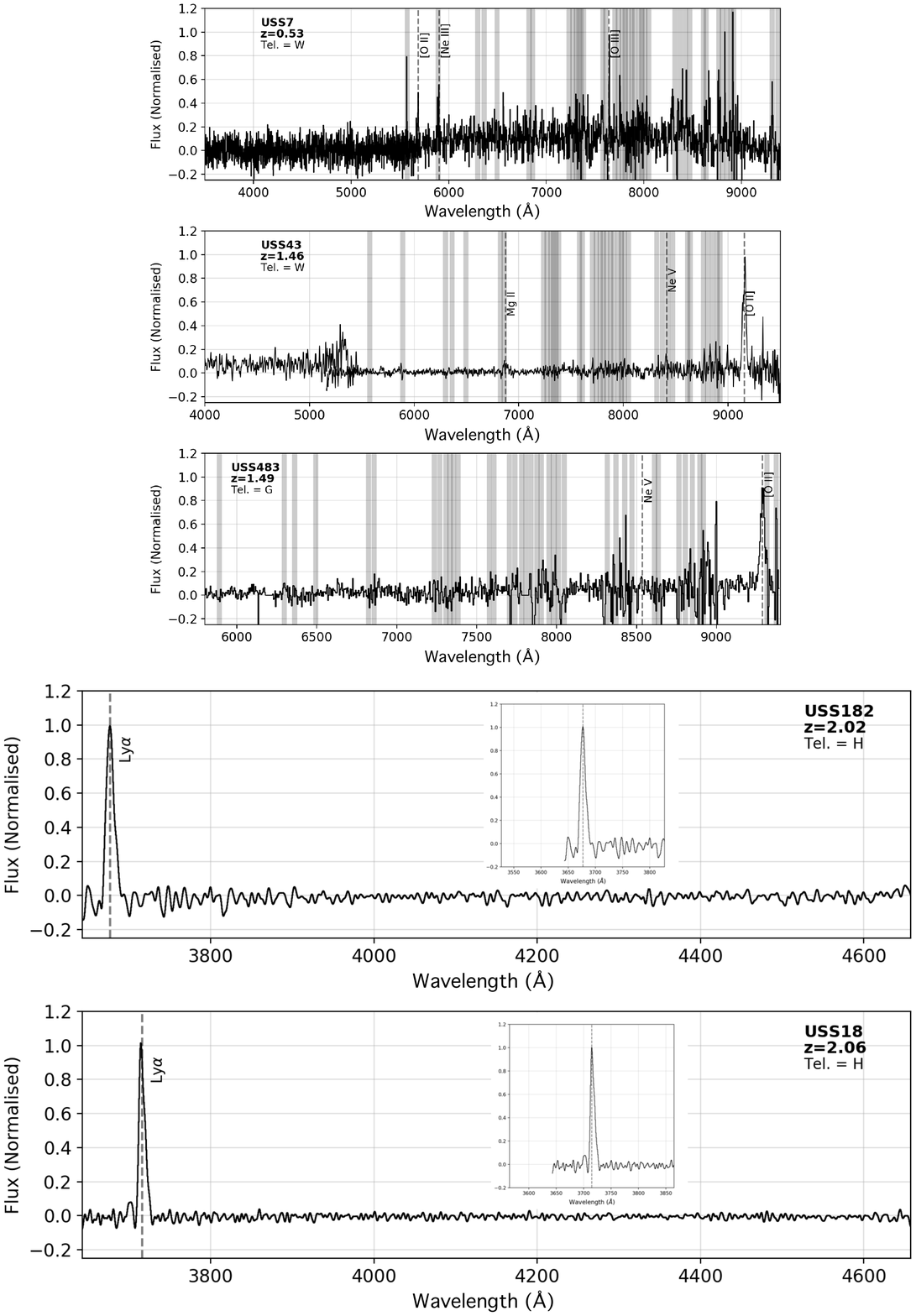}
\vspace{-30pt}

\caption{One dimensional spectra of all spectroscopically confirmed sources with the WHT, HET and Gemini. In the inset, we show a zoom-in of the \lya emission line for sources with $z>2$. The shaded boxes indicate regions of strong contamination from atmospheric sky lines. Indicated in each spectrum is the, object name, redshift and telescope code (Tel.).}
\label{fig:spectra}
\end{figure*}

\begin{figure*}
\ContinuedFloat
\captionsetup{list=off}
\includegraphics[width=0.95\textwidth]{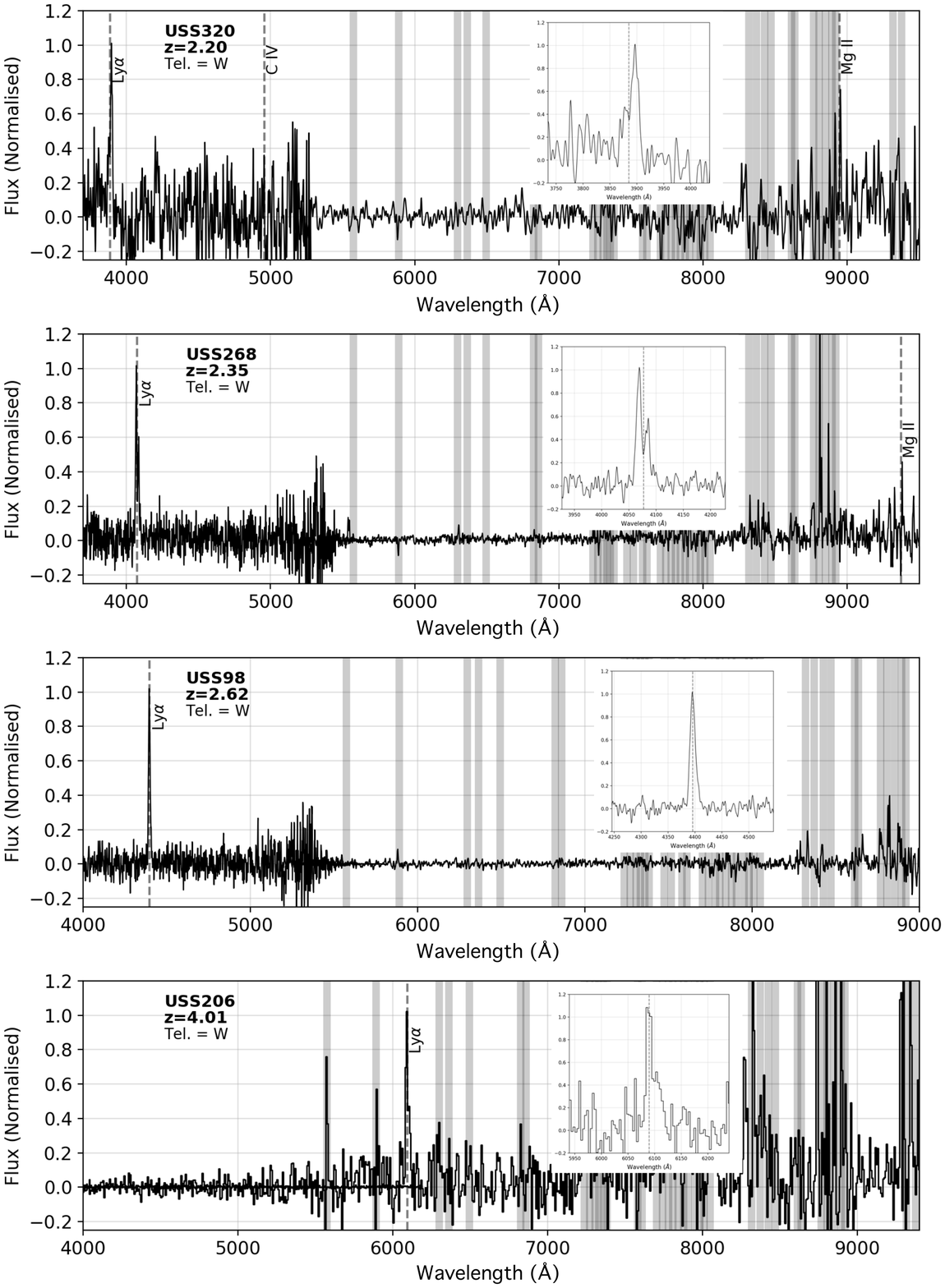}

\caption{Continued.}
\end{figure*}

\begin{figure*}
\ContinuedFloat
\captionsetup{list=off}
\includegraphics[width=0.95\textwidth]{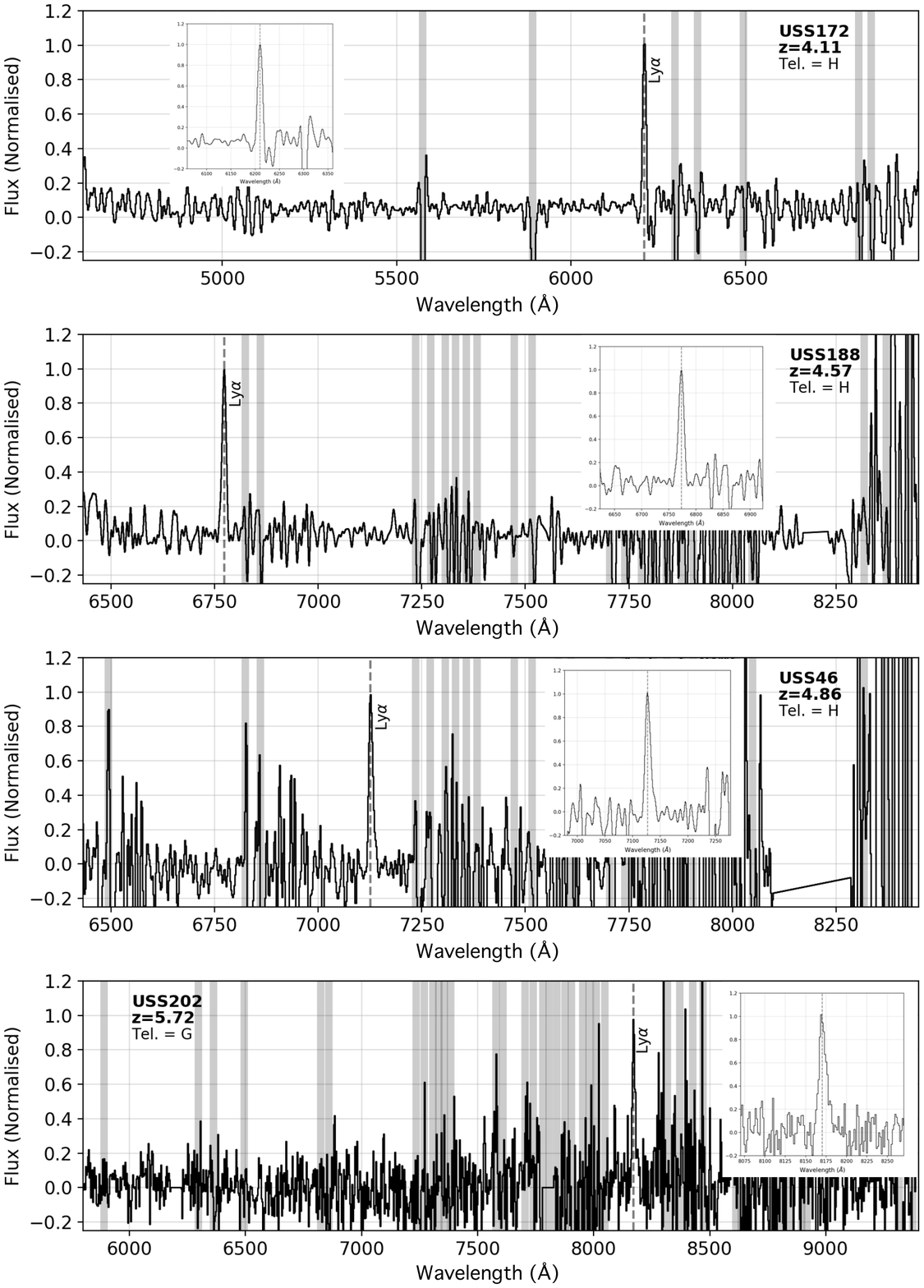}

\caption{Continued.}
\end{figure*}

\newpage
\section{K-band images}
Here we show the $K$ band images for the subset of sources that were observed at NIR wavelengths. Overlaid on top of the $K$ band image are radio contours at 1.4 GHz from \citet{sax18a}. The contours shown begin at 0.25 mJy, which is on average $3.5-5\sigma$, and are a geometric progression of $\sqrt{2}$, such that the flux density increases by a factor of 2 for every two contours.
\begin{figure*}
\centering
\includegraphics[width=.48\textwidth]{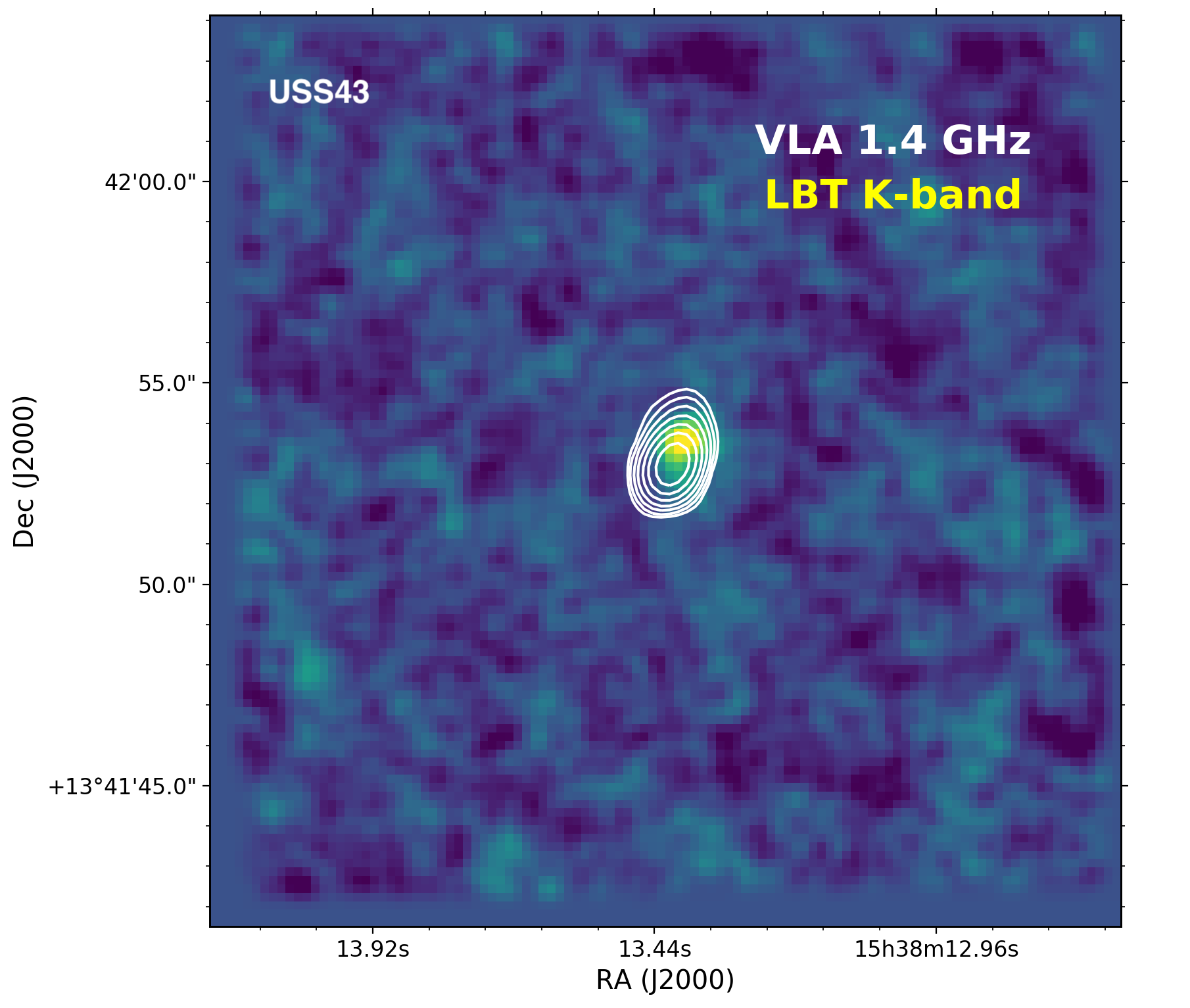}
\includegraphics[width=.48\textwidth]{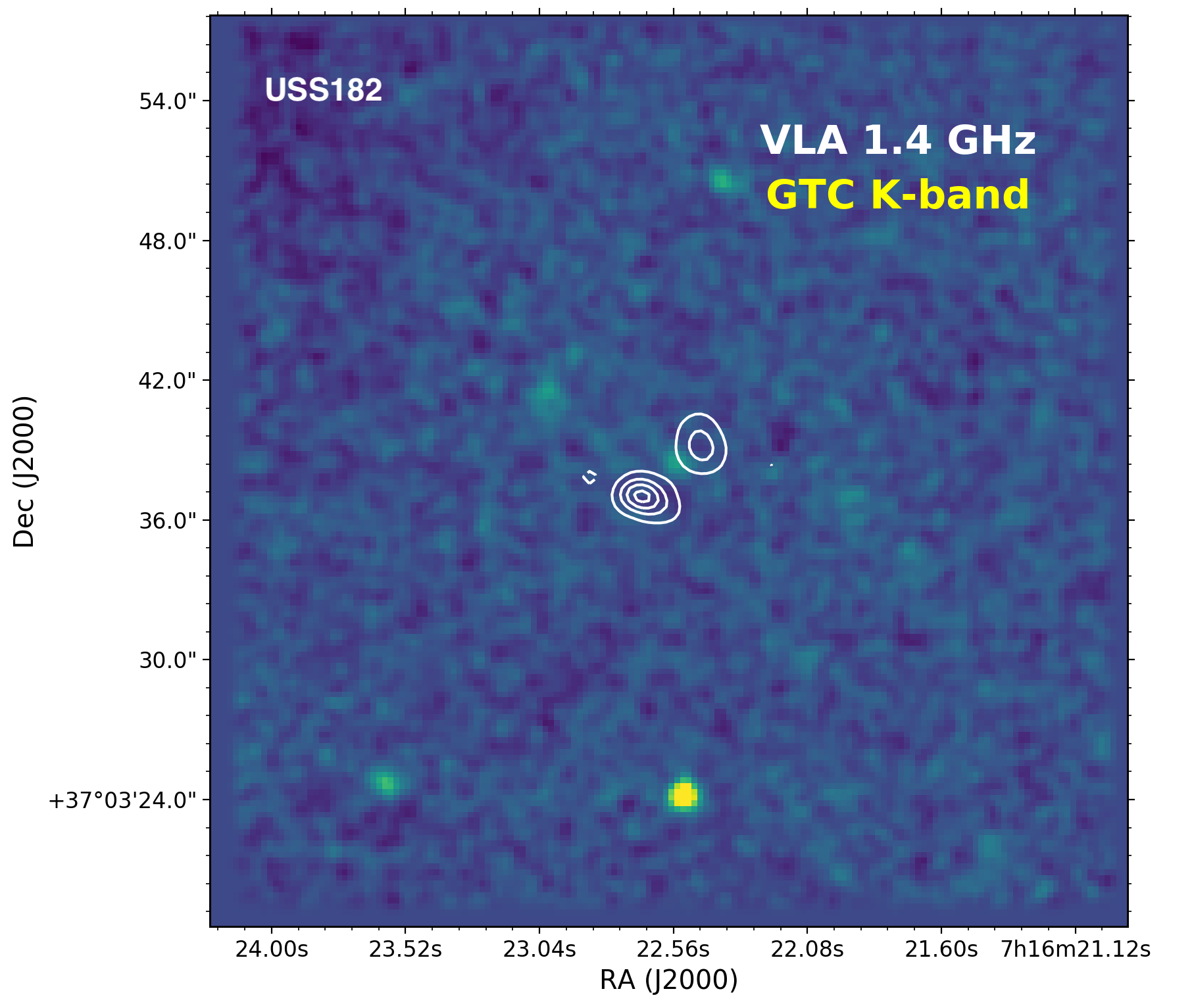}

\vspace{20pt}
\includegraphics[width=.48\textwidth]{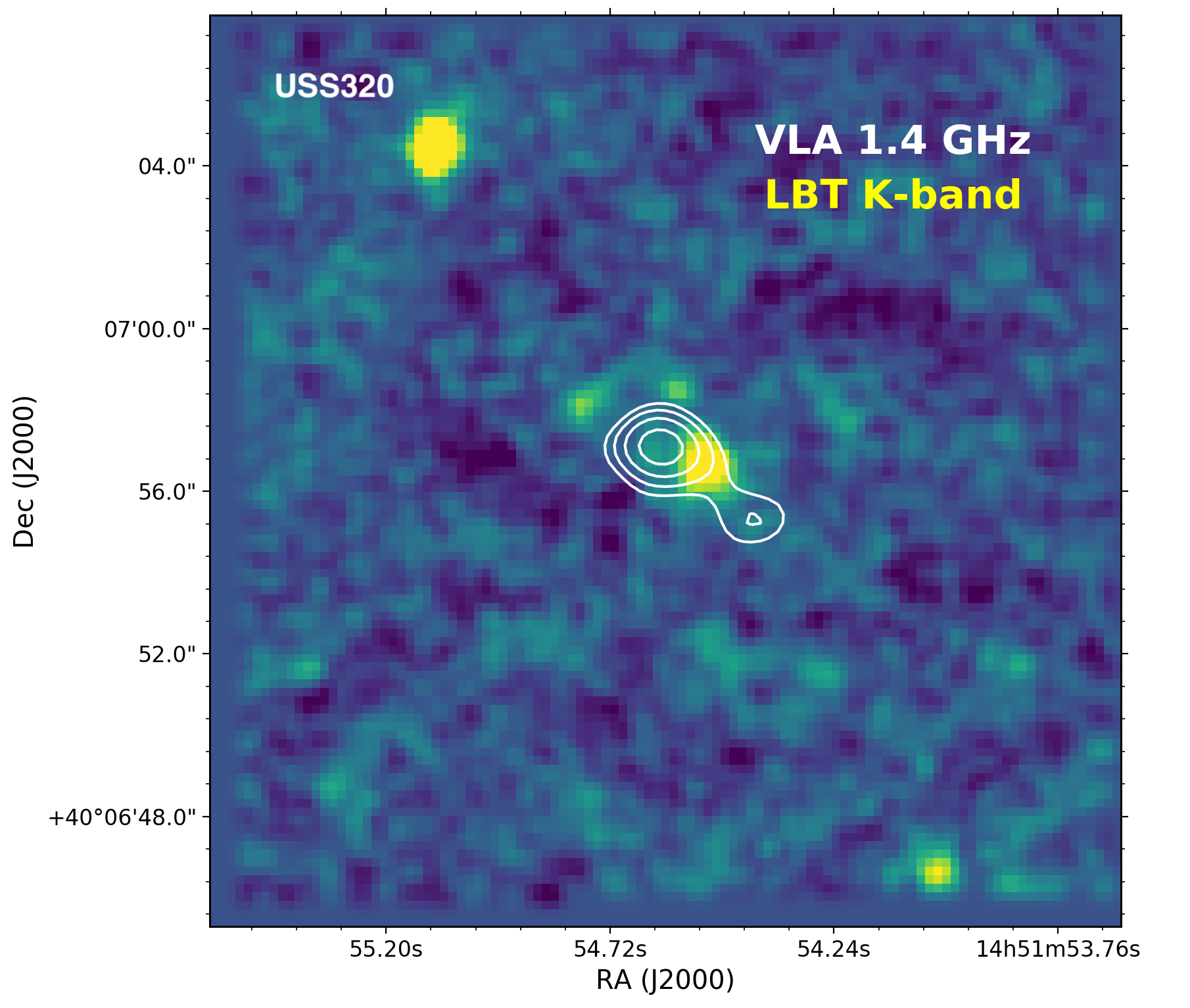}
\includegraphics[width=.48\textwidth]{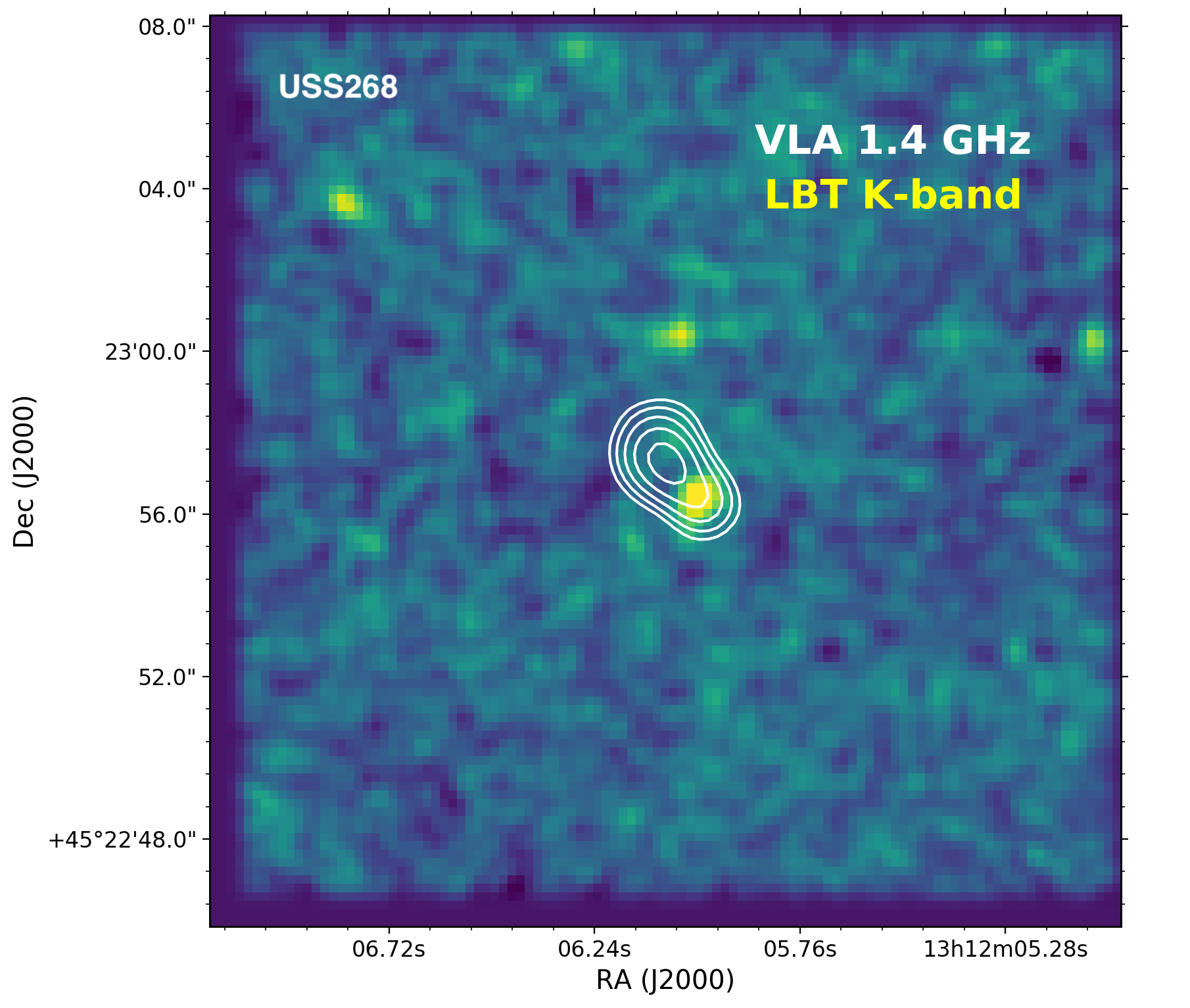}

\caption{K-band images taken from the LBT and GTC of the host galaxies of the radio sources, with radio contours at 1.4 GHz from \citet{sax18a} overlaid on top. The contours shown begin at 0.25 mJy, which is on average $3.5 - 5\sigma$, and are a geometric progression of $\sqrt{2}$, such that the flux density increases by a factor of 2 for every two contours.}
\label{fig:kband}
\end{figure*}

\begin{figure*}
\ContinuedFloat
\captionsetup{list=off}

\includegraphics[width=.48\textwidth]{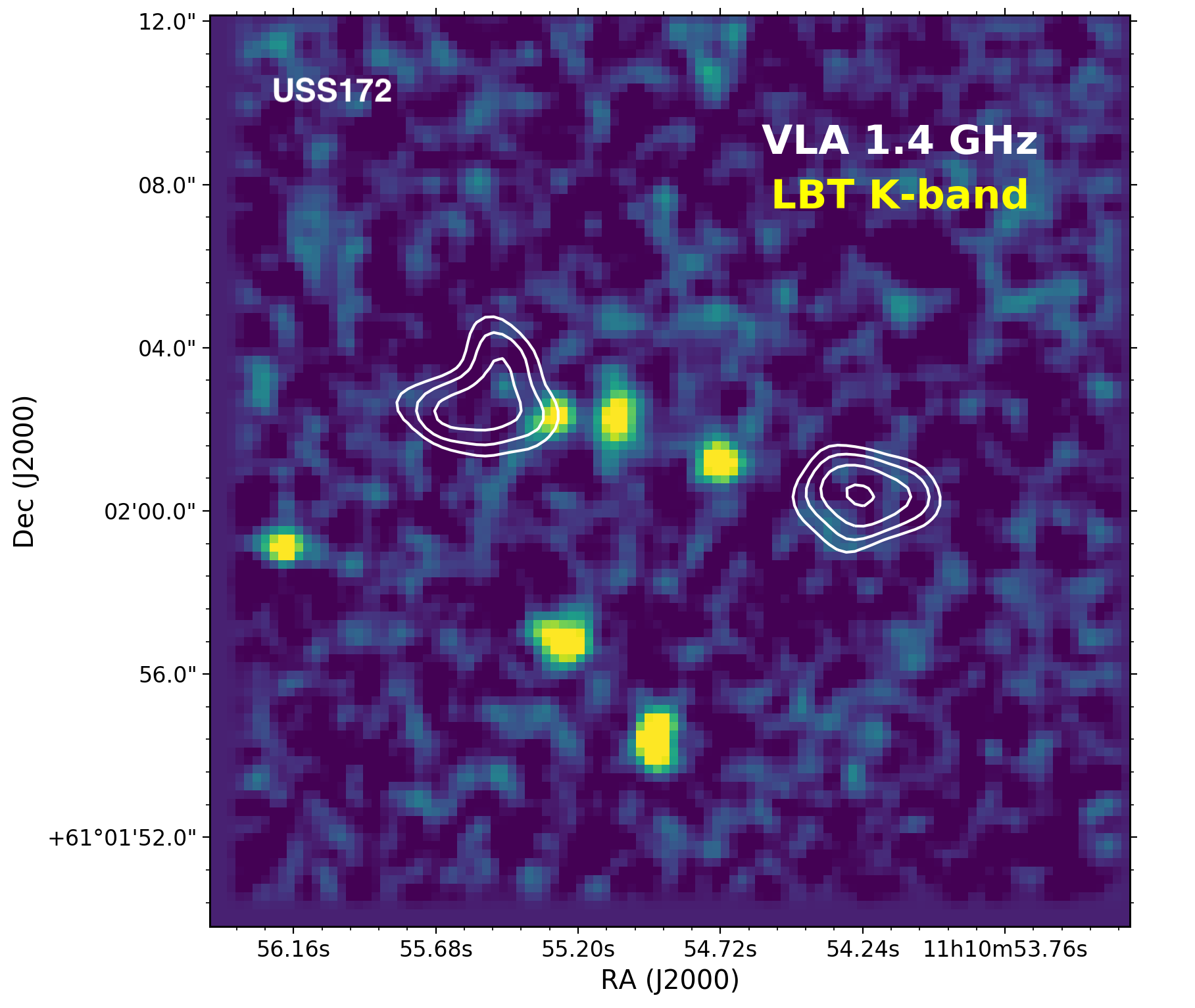}
\includegraphics[width=.48\textwidth]{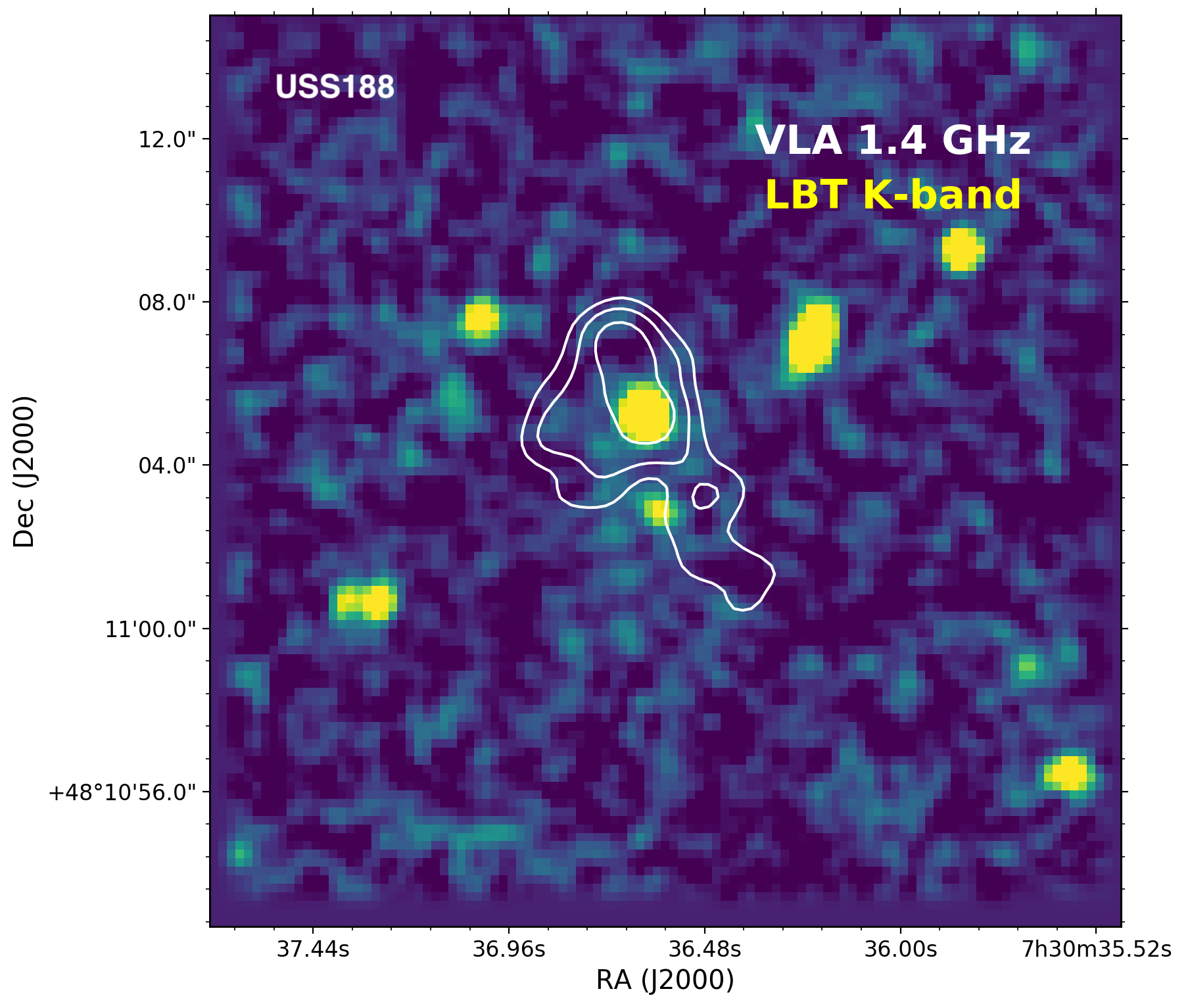}

\vspace{20pt}
\includegraphics[width=.48\textwidth]{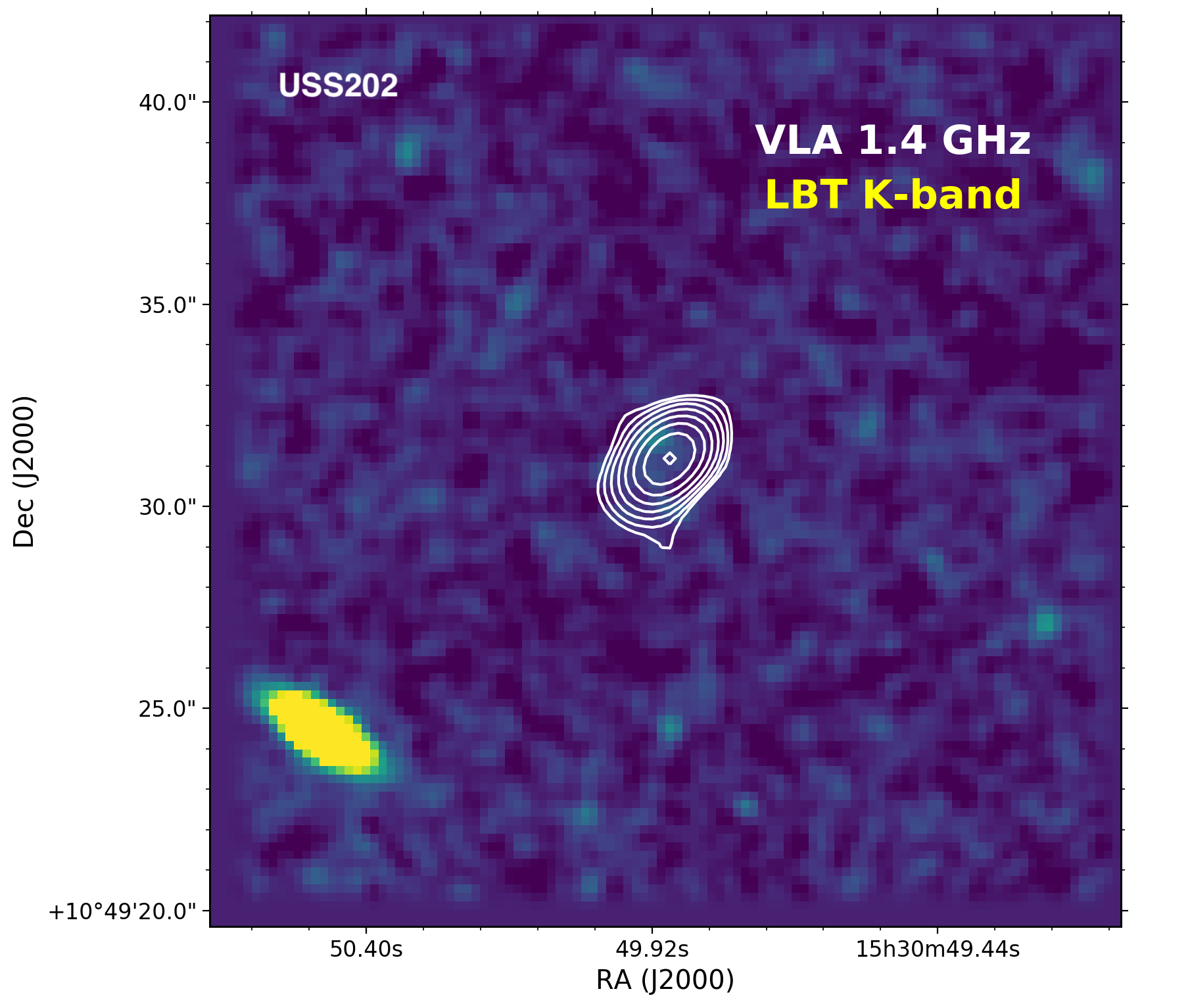}

\caption{Continued.}
\end{figure*}

% Don't change these lines
\bsp	% typesetting comment
\label{lastpage}
\end{document}